# Coarse-to-fine spatial modeling:

# A scalable, machine-learning-compatible framework

Daisuke Murakami[1,2], Alexis Comber[3], Takahiro Yoshida[4],

Narumasa Tsutsumida[5], Chris Brunsdon[6], Tomoki Nakaya[7]

[1] Department of Fundamental Statistical Mathematics, Institute of Statitsical Mathematics, Japan (E-mail: dmuraka@ism.ac.jp)

[2] Center for Urban & Real Estate Studies, Hitotsubashi University, Japan

[3] School of Geography, University of Leeds, UK

[4] Center for Spatial Information Science, the University of Tokyo, Japan

[5] Graduete School of Science & Engineering, Saitama University, Saitama, Japan

[6] National Center for Geocomputation, Maynooth University, Ireland

[7] Graduate School of Environment Studies, Tohoku University, Japan

Abstract: This study proposes coarse-to-fine spatial modeling (CFSM) as a scalable and machine learning-compatible alternative to conventional spatial process models. Unlike conventional covariance-based spatial models, CFSM represents spatial processes using a multiscale ensemble of local models. To ensure stable model training, larger-scale patterns that are easier to learn are modeled first, followed by smaller-scale patterns, with training terminated once the validation score stops improving. The training procedure, which is based on holdout validation, can be easily integrated with other machine learning algorithms, including random forests and neural networks. CFSM training is computationally efficient because it avoids explicit matrix inversion, which is a major computational bottleneck in conventional spatial Gaussian processes. Comparative Monte Carlo experiments demonstrated that the CFSM, as well as its integration with random forests, achieved superior predictive performance compared to existing models. Finally, we applied the proposed methods to an analysis of residential land prices in the Tokyo metropolitan area, Japan. The CFSM is implemented in an R package spCF (https://cran.r-project.org/web/packages/spCF/).

Keywords: Coarse-to-fine spatial modeling; local model; Gaussian process; random forest; land price; computational efficiency

# 1. Introduction

Spatial process modeling is a central topic in spatial statistics. Gaussian processes (GPs) (Williams and Rammussen, 1995; Cressie, 2015) are widely used to represent spatially dependent processes using distance-decaying covariance/correlation functions. As explained by Gelfand et al. (2010), GPs have been used for various purposes including spatial and spatiotemporal predictions, regression analysis, and uncertainty quantification.

GPs face two major challenges: scalability for large samples and limited flexibility. Regarding scalability, GPs are feasible only for small-to-moderate samples because of the need to invert an $N \times N$ covariance matrix, with computational complexity rapidly increasing on the order of $N^3$, where $N$ is the sample size. To mitigate this issue, fast GP approximations have been proposed (see Heaton, 2019; Liu et al., 2020; Hazra et al., 2025 for a review). Typical approaches include basis function approximation (see Cressie et al., 2022 for a review), which represents map patterns using $L$ (<<N) spatial basis functions, and sparse covariance/precision matrix approximation, which assumes (conditional) independence from distant samples.

Basis function approximations include fixed rank kriging (Cressie and Johannesson, 2008), generalized additive models (GAMs; Wood, 2017), and stochastic partial differential equation (SPDE) models (Bakka et al., 2018). Although they are computationally efficient, they tend to overlook small-scale spatial variations, leading to reduced modeling accuracy (Murakami et al., 2024). Sparse approximations include the stochastic local interaction model (Hristopulos, 2015; Hristopulos et al., 2021), nearest-neighbor GP (NNGP; Datta et al., 2016), and other Vecchia approximations (Katzfuss and Guiness, 2021). They tend to capture small-scale patterns accurately but often introduce noise when the true process exhibits large-scale patterns, as we will demonstrate later. The development of fast spatial process models that accurately describe both large- and small-scale map patterns remains an important research topic.

Regarding flexibility, spatial GPs are suitable for modeling continuous map patterns, but they are less suitable for modeling other patterns, including discontinuous and/or non-linear patterns. For instance, in land price modeling, spatial GPs are less suited to capturing abrupt changes in price levels between areas separated by physical or institutional barriers, such as rivers, roads, or zoning regulations. As suggested by Yoshida et al. (2023), the resulting spatial prediction accuracy of GPs is often lower than that of other machine learning algorithms such as random forest (Breiman, 2001) and gradient boosting (Freidman, 2002), particularly when many covariates/features are available. Although Gaussian spatial models have been integrated with random forests (e.g., Georanos et al., 2021), gradient boosting (Sigrist, 2022), and neural networks (e.g., Chen et al., 2020; Zammit-Mangion et al., 2022; Zhan and Datta, 2025), further research is needed to better leverage the efficiency of GPs in modeling smooth patterns by integrating them within machine learning algorithms to model non-smooth complex patterns.

A major bottleneck in achieving such integration lies in the differences between model optimization frameworks. Spatial regression models are typically estimated using likelihood-based approaches supplemented with priors in the case of Bayesian inference. In contrast, machine learning algorithms are typically optimized using validation methods. To facilitate integration, it is necessary to develop a spatial process model that can be optimized via validation.

To address these background issues related to scalability and flexibility, this study proposes a coarse-to-fine spatial modeling (CFSM) framework as a fast, validation-based alternative to conventional spatial models. This framework also facilitates seamless integration with machine learning algorithms. The remainder of this paper is organized as follows: Section 2 outlines our framework and explains the assumed spatial process. Section 3 presents the coarse-to-fine learning procedure. Section 4 compares the CFSM framework with alternative spatial models through Monte Carlo experiments, and Section 5 applies the proposed method to land price analysis. Finally, Section 6 draws some conclusions from the study.

# 2. Carse-to-fine spatial modeling (CFSM)

## 2.1. Outline

CFSM describes a multiscale spatial process over a study region $D \subset \mathbb{R}^2$. The process at site $s_i$ is defined as

$$z_{1:R}(s_i) = \sum_{r=1}^{R} z_r(s_i), \tag{1}$$

where $z_r(s_i) \sim N(\hat{z}_r(s_i), \hat{\sigma}_r^2(s_i))$ denotes the single-scale process with mean $\hat{z}_r(s_i)$ and variance $\hat{\sigma}_r^2(s_i)$ at scale *r*. The scale is characterized by the bandwidth $h_r = \delta h_{r-1}$ with a discount ratio $0 <$ $\delta < 1$. Thus, $z_1(s_i)$ is the largest-scale process, $z_2(s_i)$ is the second largest-scale process, and $z_R(s_i)$ is the $R$-th largest- or smallest-scale process. The terminal resolution $R$ is treated as an unknown, and a larger $R$ leads to a spatial process that expresses finer-scale patterns. For flexible modeling, each single-scale process $z_r(s_i)$ is defined by an aggregation (ensemble) of local models distributed over the study region. In summary, CFSM relies on the following components:

- Local models (Section 2.2), which capture local spatial patterns,
- Single-scale process $z_r(s_i)$ (Section 2.3), which aggregates the local models at each scale,
- Multiscale process $z_{1:R}(s_i)$ (Eq. 1) that integrates single-scale processes across scales.

Figure 1 summarizes the modeling procedure. Unlike conventional spatial models estimated in a single step, CFSM estimates the model stagewise ("Learning order" in Figure 1). The largest-scale process $z_1(s_i)$ is first estimated by aggregating local models associated with the largest bandwidth $h_1$, the second largest-scale process $z_2(s_i)$ is estimated similarly using the bandwidth $h_2 = \delta h_1$, and so on. This estimation continues until the $R$-th scale process $z_R(s_i)$, at which point the validation score no longer improves. They are then summed up to obtain a multiscale process $z_{1:R}(s_i)$. Since the sum of Gaussians is also a Gaussian, the aggregated process is expressed as

$$z_{1:R}(s_i) \sim N\left(\hat{z}_{1:R}(s_i), \hat{\sigma}_{1:R}^2(s_i)\right), \qquad \hat{z}_{1:R}(s_i) = \sum_{r=1}^{R} \hat{z}_r(s_i), \tag{2}$$

where $\hat{z}_{1:R}(s_i)$ is the predictive mean. The predictive variance $\hat{\sigma}_{1:R}^2(s_i)$ is evaluated through resampling as detailed in Section 3.3.

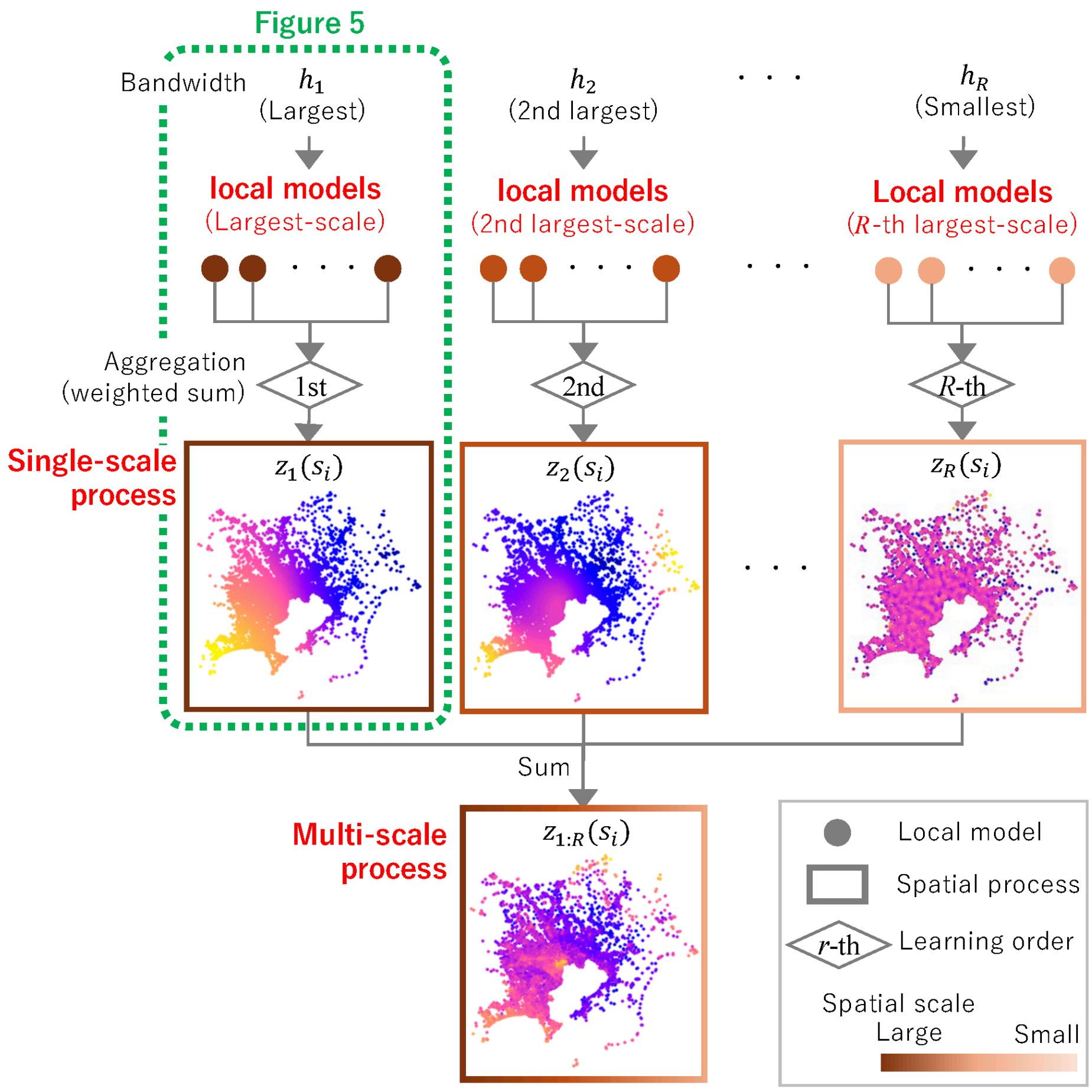

Figure 1. Procedure of CFSM (See Figure 5 for the modeling procedure of the single-scale process).

Although such coarse-to-fine sequential learning has not been explored in spatial statistics, this learning procedure offers several advantages in terms of model flexibility, computational efficiency, and expandability, which are detailed later. Below, Section 2.2 introduces the local models, and Section 2.3 aggregates them to construct the single-scale process. Section 2.4 provides a short summary.

## 2.2. Local models

At $r$-th scale, we estimate the local models indexed by $c_r \in \{1_r, \ldots, C_r\}$, assuming a pre-specified bandwidth $h_r$. Section 2.2.1 describes the local model, and Section 2.2.2 explains how these models are spatially distributed.

### 2.2.1. Model

Our local model estimates the mean $\mu_{c_r}$ and variance $v_{c_r}^2$ at the local center $s_{c_r}$ as follows:

$$z_r(s_i)|c_r \sim N\left(\mu_{c_r}, \frac{v_{c_r}^2}{w_{h_r}(d_{i,c_r})}\right), \quad \mu_{c_r} \sim N\left(0, \tau_{c_r}^2\right). \tag{3}$$

A Gaussian prior with variance $\tau_{c_r}^2$ is imposed on $\mu_{c_r}$ to regularize the estimate. The local weight $w_{h_r}(d_{i,c_r})$ is defined by a kernel decaying with the Euclidean distance $d_{i,c_r}$ from the local center, satisfying $\sum_{c_r=1}^{C_r} w_{h_r}(d_{i,c_r}) = 1$. We employ the Gaussian kernel $w_{h_r}(d_{i,c_r}) = \frac{1}{A}\exp\left(-\frac{d_{i,c_r}^2}{h_r^2}\right)$ and exponential kernel $w_{h_r}(d_{i,c_r}) = \frac{1}{A}\exp\left(-\frac{d_{i,c_r}}{h_r}\right)$, where $A$ is a normalizing constant ensuring the sum-to-one constraint $\sum_{c_r=1}^{C_r} w_{h_r}(d_{i,c_r}) = 1$. A smaller bandwidth $h_r$ assigns greater weights to nearby samples to capture small-scale patterns, whereas a larger bandwidth assigns greater weights to distant samples to capture large-scale patterns.

As detailed in Appendix 1, the predictive distribution of Eq. (3) becomes $z_r(s_i)|c_r \sim N(\hat{\mu}_{c_r}, \hat{v}_{c_r}^2(s_i))$, where

$$\hat{\mu}_{c_r} = \frac{\sum_{i=1}^{N} w_{h_r}(d_{i,c_r})\, z_r(s_i)}{\sum_{i=1}^{N} w_{h_r}(d_{i,c_r}) + \frac{v_{c_r}^2}{\tau_{c_r}^2}}, \tag{4}$$

$$\hat{v}_{c_r}^2(s_i) = \frac{\hat{v}_{c_r}^2}{\sum_{i=1}^{N} w_{h_r}(d_{i,c_r}) + \frac{\hat{v}_{c_r}^2}{\tau_{c_r}^2}} + \frac{\hat{v}_{c_r}^2}{w_{h_r}(d_{i,c_r})}, \tag{5}$$

with $\hat{v}_{c_r}^2 = \frac{\sum_{i=1}^{N} w_{h_r}(d_{i,c_r})\left(y(s_i) - \hat{\mu}_{c_r}\right)^2}{N-1}$. Figure 2 illustrates the predictive distribution for the one-dimensional case. The prediction interval, determined by the variance $\hat{v}_{c_r}^2(s_i)$, is narrow near the center $s_{c_r}$ and extremely wide in other regions, indicating that the local model has predictive power only in the vicinity of the center.

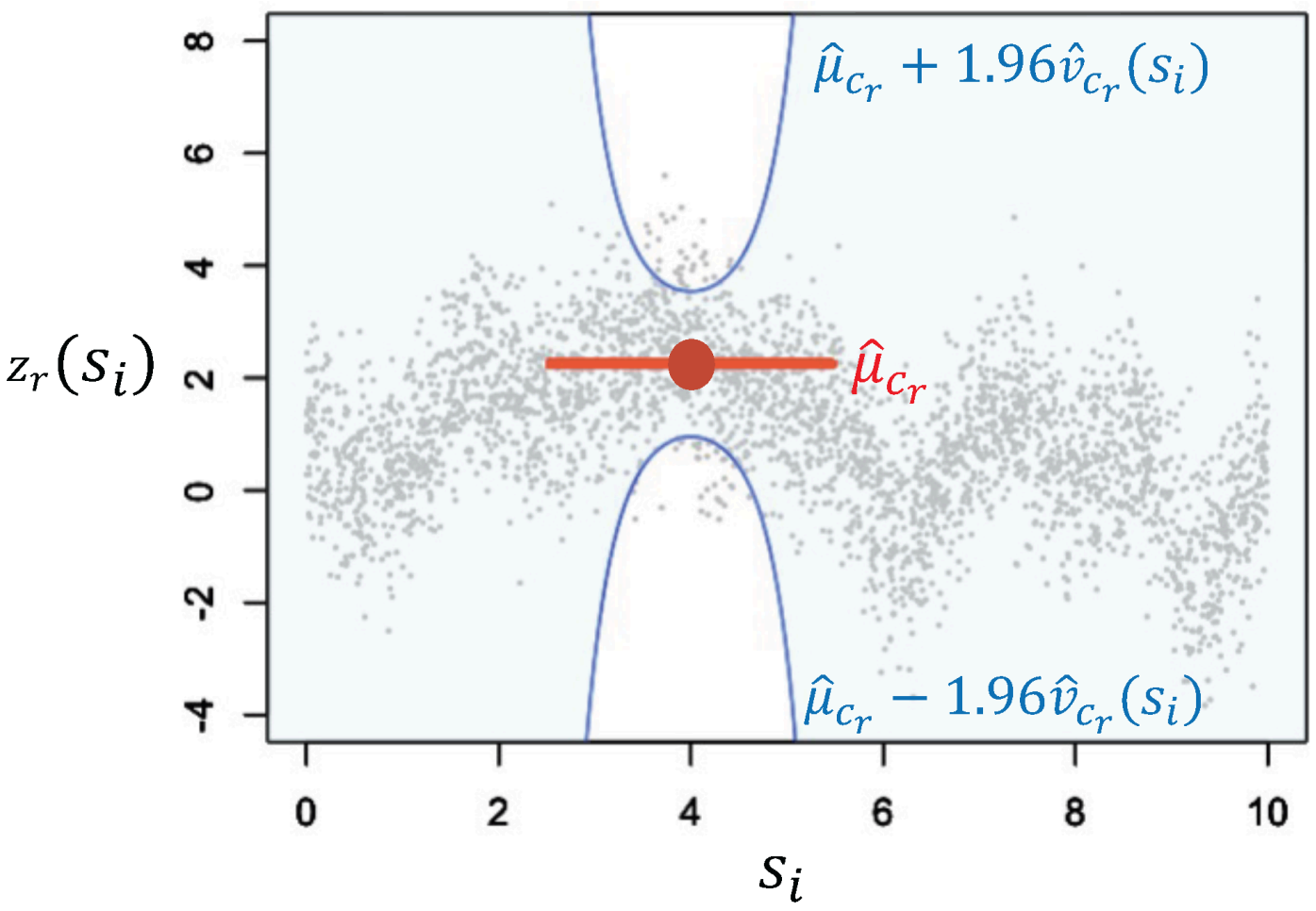

Figure 2. Local predictive mean $\hat{\mu}_{c_r}$ (red) and 95 % prediction interval (blue; $\hat{\mu}_{c_r} \pm 1.96\hat{v}_{c_r}(s_i)$) of the local model with a Gaussian kernel ($h_r = 0.8$) with center location $s_{c_r} = 4.0$. Synthetic samples (grey dots; *N*= 3,000) are generated from a moving average process, and the model is fitted.

### 2.2.2 Distributing local models

We distribute local centers $s_{1_r}, \ldots, s_{C_r}$ where local models are estimated, to collectively capture patterns across the study area. Following Kumar et al. (2012) and Murakami and Griffith (2019), who showed that *k*-means cluster centers effectively represent sample sites, we defined local centers as the sample sites closest to the *k*-means cluster centers.

The number of local centers $C_r$ is sufficient if the kernel windows (areas within a distance $h_r$ from each center) collectively cover the entire region. We set the number of centers as $C_r = round(1.5D^2/h_r^2)$, where $round(\cdot)$ denotes rounding, and $D$ is the diagonal length of the bounding square containing all sample sites. $D^2$ represents the area of the bounding square, and $h_r^2$ approximates the coverage area of each kernel. This criterion ensures that the total coverage area of $C_r$ kernels is about 1.5 times larger than the bounding square $D^2$ (i.e., $h_r^2 C_r \approx 1.5D^2$), thereby yielding a sufficiently large number of centers as shown in Figure 3. Note that alternative criteria for determining the number and placement of kernels are also possible and merit further investigation in future work.

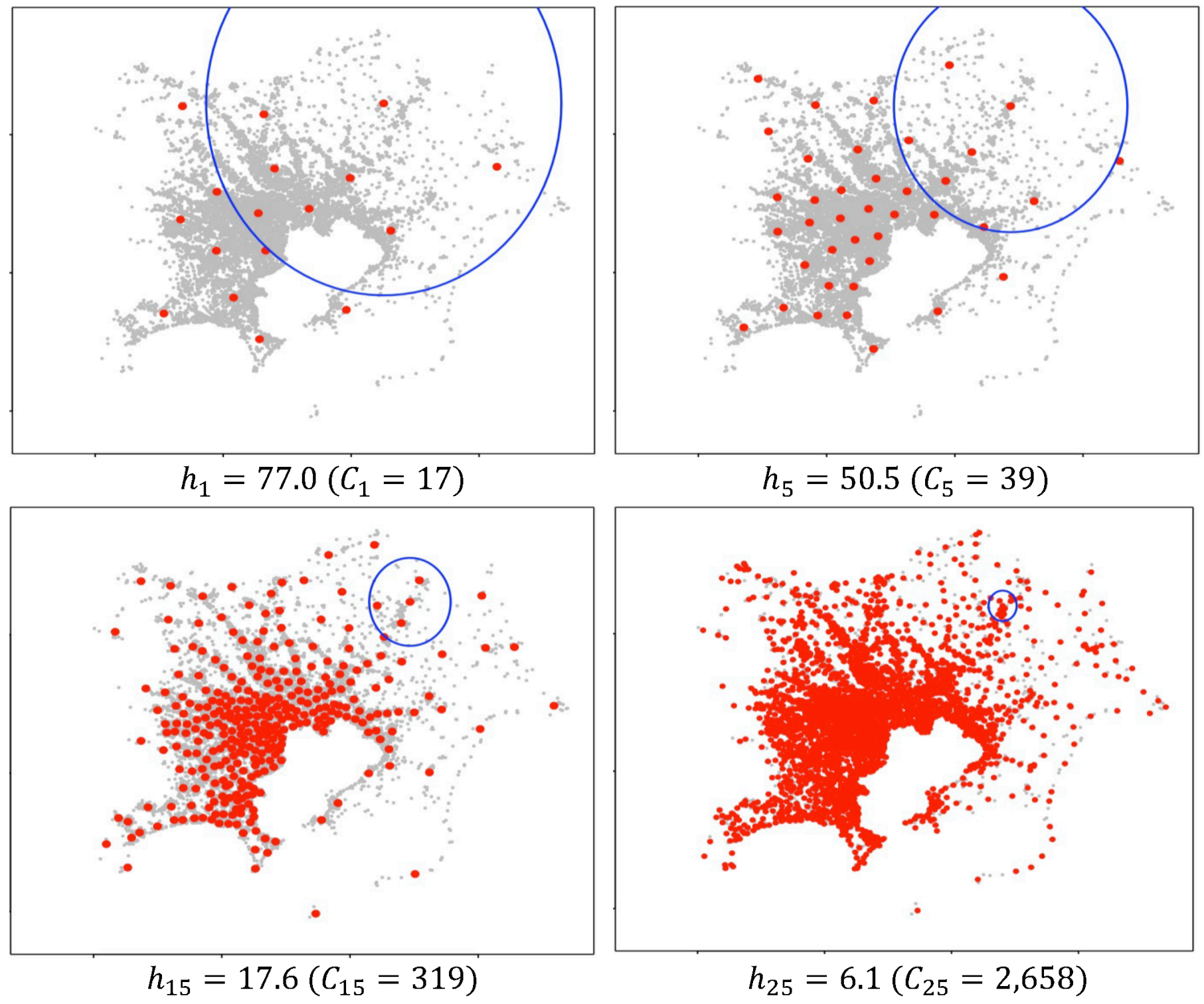

Figure 3. $C_r$ local centers (red) for four bandwidth values. The centers are selected from land price assessment sites (grey; see Section 5). Blue circles indicate kernel window with radius $h_r$. The large kernel window in each panel suggests that the kernels in each scale collectively cover the study area.

## 2.3. Single-scale spatial process

In this section, the local models are aggregated to construct a single-scale process. In distribution form, the $c_r$-th local model can be expressed as [1]

$$p(z_r(s_i)|c_r) \propto p\left(z_r(s_i)\middle|\hat{\mu}_{c_r}, \hat{v}_{c_r}^2\right)^{w_{h_r}(d_{i,c_r})}. \tag{6}$$

[1] The probability density function of $z_r(s_i)|c_r \sim N\left(\mu_{c_r}, \hat{v}_{c_r}^2 / w_{h_r}(d_{i,c_r})\right)$ is $p(z_r(s_i)|c_r) = \frac{w_{h_r}(d_{i,c_r})}{\sqrt{2\pi}\hat{v}_{c_r}} \exp\left(-\frac{w_{h_r}(d_{i,c_r})}{2\hat{v}_{c_r}^2}\left(y(s_i) - \hat{\mu}_{c_r}\right)^2\right) = w_{h_r}(d_{i,c_r}) \frac{1}{\sqrt{2\pi}\hat{v}_{c_r}} \exp\left(-\frac{1}{2\hat{v}_{c_r}^2}\left(y(s_i) - \hat{\mu}_{c_r}\right)^2\right)^{w_{h_r}(d_{i,c_r})} = w_{h_r}(d_{i,c_r}) p\left(y(s_i)\middle|\hat{\mu}_{c_r}, \hat{v}_{c_r}^2\right)^{w_{h_r}(d_{i,c_r})} \propto p\left(y(s_i)\middle|\hat{\mu}_{c_r}, \hat{v}_{c_r}^2\right)^{w_{h_r}(d_{i,c_r})}$.

$p\big(z_r(s_i)\big|\hat{\mu}_{c_r}, \hat{v}_{c_r}^2\big)$ is the probability density function of $z_r(s_i) \sim N(\hat{\mu}_{c_r}, \hat{v}_{c_r}^2)$, where $\mu_{c_r} \sim N(0, \tau_{c_r}^2)$ present in Eq. (3) disappears because $\mu_{c_r}$ is fixed by the estimate $\hat{\mu}_{c_r}$.

A single-scale process is defined as the product of local model distributions:

$$p(z_r(s_i)|1_r, \dots, C_r) \propto \prod_{c_r=1_r}^{C_r} p\big(z_r(s_i)\big|\hat{\mu}_{c_r}, \hat{v}_{c_r}^2\big)^{w_{h_r}(d_{i,c_r})}. \quad (7)$$

where $\sum_{c_r=1}^{C_r} w_{h_r}\big(d_{i,c_r}\big) = 1$ as assumed in Section 2.2.1. This model aggregation, known as the generalized product-of-experts (Cao and Fleet, 2014), was adopted in this study because it provides an optimal model that is closest to each local model in terms of the weighted Kullback–Leibler divergence (Cao and Fleet, 2015). In other words, the aggregated model preserves the local patterns captured by each local model as much as possible. The accuracy of the aggregated model has been empirically demonstrated (Cohen et al., 2020; Murakami et al., 2024).

As shown in Appendix 2, Eq. (7) can be expressed as a single-scale spatial process.

$$z_r(s_i) \sim N(\hat{z}_r(s_i), \hat{\sigma}_r^2(s_i)), \quad (8)$$

where

$$\hat{z}_r(s_i) = \frac{\sum_{c_r=1}^{C_r} w_{c_r}(s_i)\hat{\mu}_{c_r}}{\sum_{c_r=1}^{C_r} w_{c_r}(s_i)}, \quad (9)$$

$$\hat{\sigma}_r^2(s_i) = \frac{1}{\sum_{c_r=1}^{C_r} w_{c_r}(s_i)}. \quad (10)$$

The predicted mean (Eq. 9) averages the local means $\hat{\mu}_{c_r}$ considering the weight $w_{c_r}(s_i) = \frac{w_r(d_{i,c_r})}{\hat{v}_{c_r}^2}$, emphasizing nearby accurate local models.

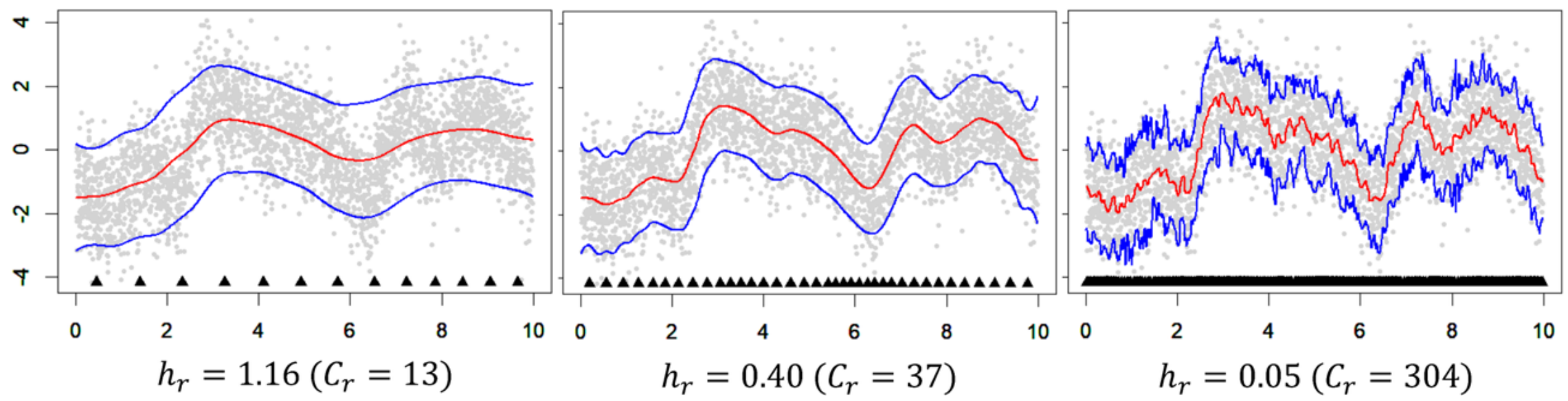

Figure 4. Predictive mean (red; $\hat{z}_r(s_i)$) and 95% confidence interval (blue; $\hat{z}_r(s_i) \pm 1.96\hat{\sigma}_r(s_i)$) of the single-scale processes with $h_r \in \{1.16, 0.40, 0.05\}$. These processes are obtained by fitting $C_r$ local models for the 3,000 samples generated from moving average processes and aggregating them using Eqs. (9) and (10). Black triangles at the bottom of each panel indicate the $C_r$ local centers.

Figure 4 shows the single-scale processes for the one-dimensional case. As shown in the figure, the process underfits when $h_r$ is too large, and overfits when $h_r$ is too small. Section 3 develops an algorithm for estimating multiscale spatial processes (Eq. 1), which synthesizes scale-wise processes while avoiding underfitting and overfitting.

## 2.4. Summary

Figure 5 summarizes the procedure for single-scale process modeling. Given a bandwidth $h_r$, the $C_r$ local centers are distributed across the study area. For each center, the local mean $\mu_{c_r}$ and variance $v_{c_r}^2$ are estimated using Eqs. (4) and (5), respectively. They are then aggregated using Eqs.

(9) and (10), respectively, to construct a single-scale process. The processes at each scale were synthesized to construct the multiscale process $z_{1:R}(s_i)$ (Eq. 1), as detailed in Section 3.

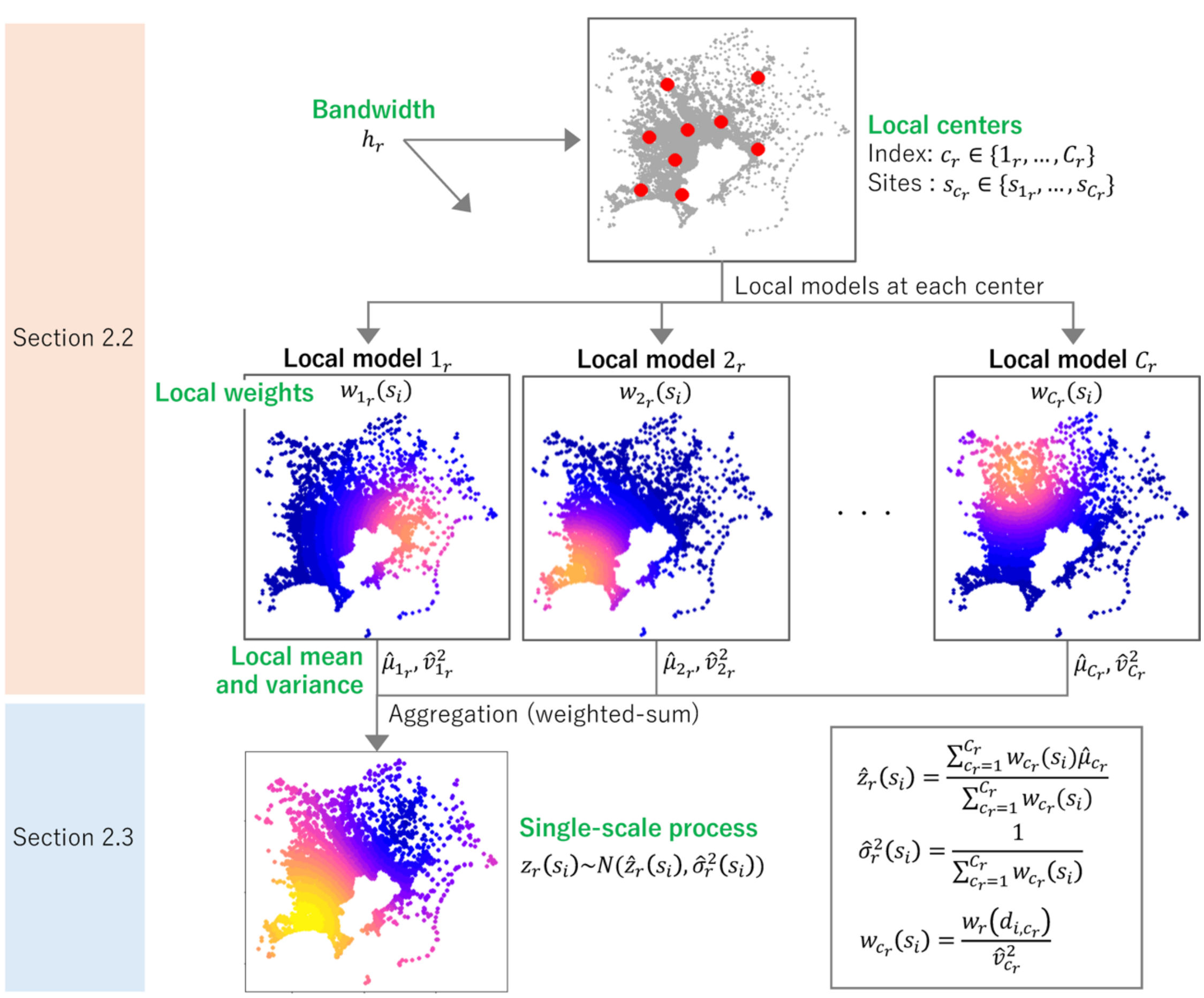

Figure 5. Procedure for modeling the single-scale spatial process at scale *r*.

# 3. Learning algorithms

Sections 3.1 and 3.2 explain our learning algorithms for linear and non-linear models with the spatial process $z_{1:R}(s_i)$, respectively, focusing on spatial prediction. Section 3.3 quantifies uncertainty. Section 3.4 compares the properties of the proposed method with those of conventional methods. For notational simplicity, we use the matrix expression in this section.

## 3.1. Linear spatial modeling

### 3.1.1. Model

We first consider the following linear spatial model:

$$\mathbf{y} = \mathbf{X}\boldsymbol{\beta} + \mathbf{z}_{1:R} + \mathbf{e}, \quad \mathbf{z}_{1:R} = \sum_{r=1}^{R} \mathbf{z}_r, \qquad \mathbf{e} \sim N(\mathbf{0}, \sigma^2 \mathbf{I}), \tag{11}$$

where $\mathbf{z}_{1:R} = [z_{1:R}(s_1), \cdots, z_{1:R}(s_N)]'$, $\mathbf{z}_r = [z_r(s_1), \cdots, z_r(s_N)]'$, and $z_r(s_i) \sim N\left(\hat{z}_r(s_i), \hat{\sigma}_r^2(s_i)\right)$. $\mathbf{y}$ is a response vector, $\mathbf{X}$ is a matrix of $K$ covariates, $\boldsymbol{\beta}$ is a coefficient vector, $\mathbf{0}$ is a vector of zeros, and $\mathbf{I}$ is an identity matrix. This model consists of the trend term $\mathbf{X}\boldsymbol{\beta}$, spatial process $\mathbf{z}_{1:R}$, and noise $\mathbf{e}$ with variance $\sigma^2$.

Similarly, the following model is assumed at the prediction site $s_0$:

$$y(s_0) = \mathbf{x}(s_0)\boldsymbol{\beta} + z_{1:R}(s_0) + e(s_0), \quad z_{1:R}(s_0) = \sum_{r=1}^{R} z_r(s_0), \qquad e(s_0) \sim N(0, \sigma^2), \tag{12}$$

where $z_r(s_0) \sim N\left(\hat{z}_r(s_0), \hat{\sigma}_r^2(s_0)\right)$, and $\mathbf{x}(s_0)$ is a covariate vector. A primal objective of our model was to predict $y(s_0)$, which was assumed to be unknown.

### 3.1.2. Learning algorithm

Simultaneously optimizing $\mathbf{z}_1, \ldots, \mathbf{z}_R$ (and $z_1(s_0), \ldots, z_R(s_0)$) is extremely challenging because of their identifiability issues and computational inefficiency. Therefore, we adopted progressive learning (see Soviany et al., 2022), in which simpler and easier-to-learn patterns are estimated first, followed by progressively more complex patterns. In our case, the appropriate learning order is $\mathbf{z}_1, \ldots, \mathbf{z}_R$ because $\mathbf{z}_r$ corresponding to a larger *r* relies on a greater number of local models to capture the localized patterns. This approach avoids the computational burden of simultaneous optimization and stabilizes estimation by prioritizing simpler processes. Appendix 3 confirms that our assumed learning order achieves the highest predictive accuracy among alternatives.

At each step of progressive learning, holdout validation (HV) was performed to assess model accuracy. In HV, data are randomly split into $100\varphi$ % training and $100(1-\varphi)$ % validation samples, with $\varphi$ set to 0.75. In other words, 75 % of the samples were used for model estimation, and the remaining 25 % for accuracy evaluation. The response vector is partitioned as $\mathbf{y} = \begin{bmatrix} \mathbf{y}_{(t)} \\ \mathbf{y}_{(v)} \end{bmatrix}$, where $\mathbf{y}_{(t)}$ and $\mathbf{y}_{(v)}$ denote sub-vectors corresponding to the training and validation samples, respectively. $\mathbf{X} = \begin{bmatrix} \mathbf{X}_{(t)} \\ \mathbf{X}_{(v)} \end{bmatrix}$, $\mathbf{z}_R = \begin{bmatrix} \mathbf{z}_{R(t)} \\ \mathbf{z}_{R(v)} \end{bmatrix}$, and sample sites $\mathbf{s} = [s_1, \ldots, s_N]' = \begin{bmatrix} \mathbf{s}_{(t)} \\ \mathbf{s}_{(v)} \end{bmatrix}$ were partitioned similarly.

The learning procedure for spatial prediction is summarized as a pseudo-code in Figure 6. The procedure is explained by referring to line *x* of the pseudo-code (Figure 6) as "(line *x*)." We perform two HVs. The first HV (lines 4–33) sequentially optimizes the single-scale processes $\mathbf{z}_1, \ldots, \mathbf{z}_R$ (learning order 1 to $R$ in Figure 1), and the second HV (lines 34–35) adjusts the estimated processes to mitigate potential bias (learning order $R$ + 1 in Figure 1). Both HVs minimize the sum of squared error (SSE) for validation samples, defined as $SSE_R = \left\|\mathbf{y}_{(v)} - \mathbf{X}_{(v)}\boldsymbol{\beta} + \hat{\mathbf{z}}_{1:R(v)}\right\|^2$.

|  |  |  |
| --- | --- | --- |
| 1 | Data: $\mathbf{y} = [\mathbf{y}'_{(t)}, \mathbf{y}'_{(v)}]'$, $\mathbf{X} = [\mathbf{X}'_{(t)}, \mathbf{X}'_{(v)}]'$, $\mathbf{s} = [\mathbf{s}'_{(t)}, \mathbf{s}'_{(v)}]'$, $\mathbf{x}(s_0)$, | |
| 2 | Parameters: $R = 1$, $Q = 0$, $Q_{\max} = 5$, $\delta = 0.9$, $B$=200, $\tau_R^2 = SSE_{R-1} = \infty$, $\hat{\mathbf{z}}_{1:R-1} = \mathbf{0}$, $h_R, D$ | |
| 3 | **Algorithm** | |
| 4 | While $Q < Q_{\max}$ | # $Q$ counts consecutive failures to improve $SSE_R$ |
| 5 | ############ **Update regression coefficients** ############ | |
| 6 | $\widehat{\boldsymbol{\beta}} \leftarrow (\mathbf{X}'\mathbf{X})^{-1}\mathbf{X}'(\mathbf{y} - \hat{\mathbf{z}}_{1:R-1})$ | |
| 7 | $\hat{\mathbf{e}}_R \leftarrow \mathbf{y} - \mathbf{X}\widehat{\boldsymbol{\beta}} - \hat{\mathbf{z}}_{1:R-1}$ | |
| 8 | ############ **Update spatial process** ################## | |
| 9 | #### Modeling single-scale process at the $R$-th scale | |
| 10 | $C_R \leftarrow \text{round}(1.5D^2/h_R^2)$ | # Number of local centers |
| 11 | $\mathbf{s}_R = [s_{1_R}, \ldots, s_{C_R}] \leftarrow \text{kmeans.centers}(\mathbf{s}_{(t)}, C_R)$ | # Distribute $C_R$ local centers |
| 12 | $\widehat{\boldsymbol{\mu}}_R, \hat{\mathbf{v}}_R^2 \leftarrow$ (empty vector of length $C_R$) | |
| 13 | $\hat{\mathbf{v}}_R^2 \leftarrow$ (empty vector of length $C_R$) | |
| 14 | For $c_R = 1_R, \ldots, C_R$ | # Local modeling for each center |
| 15 | $\widehat{\boldsymbol{\mu}}_R[c_R], \hat{\mathbf{v}}_R^2[c_R] \leftarrow \text{local.model}(s_{c_R}, \mathbf{s}_{(t)}, \hat{\mathbf{e}}_{R(t)}, h_{\mathrm{R}}, \tau_R^2)$ | # Local mean and variance (Eqs.3,4) |
| 16 | End For | |
| 17 | $\hat{\mathbf{z}}_{R(v)}, \widehat{\boldsymbol{\sigma}}_{R(v)}^2 \leftarrow \text{ss.process}(\mathbf{s}_{(v)}, \mathbf{s}_R, \widehat{\boldsymbol{\mu}}_R, \hat{\mathbf{v}}_R^2)$ | # Process at validation sites $\mathbf{s}_{(v)}$ (Eqs.7,8) |
| 18 | #### Accept $\hat{\mathbf{z}}_{R(v)}$ if it reduces the validation SSE | |
| 19 | $SSE_R \leftarrow \lVert \mathbf{y}_{(v)} - \mathbf{X}_{(v)}\widehat{\boldsymbol{\beta}} - \hat{\mathbf{z}}_{1:R-1(v)} - \hat{\mathbf{z}}_{R(v)} \rVert^2$ | # Validation SSE given $\hat{\mathbf{z}}_{R(v)}$ |
| 20 | If $SSE_R < SSE_{R-1}$ | |
| 21 | $\hat{\mathbf{z}}_{R(t)}, \widehat{\boldsymbol{\sigma}}_{R(t)}^2 \leftarrow \text{ss.process}(\mathbf{s}_{(t)}, \mathbf{s}_R, \widehat{\boldsymbol{\mu}}_R, \hat{\mathbf{v}}_R^2)$ | # Process at training sites $\mathbf{s}_{(t)}$ (Eqs.7,8) |
| 22 | $\hat{z}_R(s_0), \hat{\sigma}_R^2(s_0) \leftarrow \text{ss.process}(s_0, \mathbf{s}_R, \widehat{\boldsymbol{\mu}}_R, \hat{\mathbf{v}}_R^2)$ | # Process at prediction site $s_0$ (Eqs.7,8) |
| 23 | $\hat{\mathbf{z}}_{1:R} \leftarrow \hat{\mathbf{z}}_{1:R-1} + [\hat{\mathbf{z}}'_{R(t)}, \hat{\mathbf{z}}'_{R(v)}]'$ | # Updated multiscale process (Eq. 1) |
| 24 | $Q \leftarrow 0$ | |
| 25 | Else | |
| 26 | $SSE_R \leftarrow SSE_{R-1}$ | |
| 27 | $\hat{\mathbf{z}}_{1:R} \leftarrow \hat{\mathbf{z}}_{1:R-1}$ | |
| 28 | $Q \leftarrow Q + 1$ | |
| 29 | End If | |
| 30 | $h_{R+1} \leftarrow \delta h_R$ | # Update bandwidth |
| 31 | $\tau_{R+1}^2 \leftarrow Var[\mu_{1_R}, \ldots, \mu_{C_R}]$ | # Update prior variance |
| 32 | $R \leftarrow R + 1$ | |
| 33 | End While | |
| 34 | $\hat{\theta}_1, \hat{\theta}_2 \leftarrow \text{argmin}_{\theta_1, \theta_2} \lVert \mathbf{y}_{(v)} - \mathbf{X}_{(v)}\widehat{\boldsymbol{\beta}} - \sum_{r=1}^{R} \alpha_r \hat{\mathbf{z}}_{r(v)} \rVert^2$ where $\alpha_r = \theta_1 exp(-\theta_2 h_r)$ | |
| 35 | $E[y(s_0)] \leftarrow \mathbf{x}(s_0)\widehat{\boldsymbol{\beta}} + \sum_{r=1}^{R} \hat{\alpha}_r \hat{z}_r(s_0)$ where $\hat{\alpha}_r = \hat{\theta}_1 exp(-\hat{\theta}_2 h_r)$ | # Predictive mean |
| 36 | $V[y(s_0)] \leftarrow \text{boot}(\mathbf{x}(s_0), \hat{z}_1(s_0), \ldots, \hat{z}_R(s_0), \hat{\sigma}_1^2(s_0), \ldots, \hat{\sigma}_R^2(s_0), \hat{\alpha}_r, B)$ | # Predictive variance |

Figure 6. Pseudo-code of CFSM-based spatial prediction. $\text{kmeans.centers}(\mathbf{s}_{(t)}, C_R)$ applies the $k$-means clustering of the sites $\mathbf{s}_{(t)}$ to extract $C_R$ local centers; $\text{local.model}(s_{c_R}, \mathbf{s}_{(t)}, \hat{\mathbf{e}}_{R(t)}, h_{\mathrm{R}}, \tau_R^2)$ fits the local model at $s_{c_R}$ on $\hat{\mathbf{e}}_{R(t)}$ to evaluate the local mean and variance; $\text{ss.process}(\mathbf{s}, \mathbf{s}_R, \widehat{\boldsymbol{\mu}}_R, \hat{\mathbf{v}}_R^2)$ evaluates the mean $\hat{\mathbf{z}}_R$ and variance $\widehat{\boldsymbol{\sigma}}_R^2$ of the single-scale process at sites $\mathbf{s}$; the boot function evaluates variance through bootstrap resampling (see Section 3.3).

### 3.1.3. First HV

The first HV estimates $\boldsymbol{\beta}$ and $\boldsymbol{z}_1, \dots, \boldsymbol{z}_R$. Given $\hat{\mathbf{z}}_{1:R-1}$, the linear spatial model (Eq. 11) becomes $\mathbf{y} = \mathbf{X}\boldsymbol{\beta} + \hat{\mathbf{z}}_{1:R-1} + \mathbf{e}$, $\mathbf{e} \sim N(\mathbf{0}, \sigma^2\mathbf{I})$. Therefore, $\boldsymbol{\beta}$ is estimated using the ordinary least squares method as $\hat{\boldsymbol{\beta}} = (\mathbf{X}'\mathbf{X})^{-1}\mathbf{X}'(\mathbf{y} - \hat{\mathbf{z}}_{1:R})$ (line 6), updating the residual vector as $\hat{\mathbf{e}}_R = \mathbf{y} - \mathbf{X}\hat{\boldsymbol{\beta}} - \hat{\mathbf{z}}_{1:R-1}$ (line 7).

Given $\hat{\mathbf{z}}_{1:R-1}$ and $\hat{\boldsymbol{\beta}}$, the *R*-th single-scale process $\mathbf{z}_R$ is newly introduced as $\mathbf{y} = \mathbf{X}\hat{\boldsymbol{\beta}} + \hat{\mathbf{z}}_{1:R-1} + \mathbf{z}_R + \mathbf{e}$. To estimate $\mathbf{z}_R$, $C_R$ local centers are determined using the *k*-means centers (line 11). The local models are fitted on $\hat{\mathbf{e}}_R$ at each center (lines 14–16) to estimate the local means $\hat{\boldsymbol{\mu}}_R = [\hat{\mu}_{1_R}, \dots, \hat{\mu}_{C_R}]'$ and variances $\hat{\mathbf{v}}_R^2 = [\hat{v}_{1_R}^2, \dots, \hat{v}_{C_R}^2]'$. They are aggregated using Eq. (9) to obtain the predictive means $\hat{\mathbf{z}}_R = \begin{bmatrix} \hat{\mathbf{z}}_{R(t)} \\ \hat{\mathbf{z}}_{R(v)} \end{bmatrix}$ and $\hat{z}_R(s_0)$ of the single-scale process (lines 17, 21, and 22). If $SSE_R = \left\| \mathbf{y}_{(v)} - \mathbf{X}_{(v)}\hat{\boldsymbol{\beta}} - \hat{\mathbf{z}}_{1:R-1(v)} - \hat{\mathbf{z}}_{R(v)} \right\|^2$ reduces, the process estimate is updated as $\hat{\mathbf{z}}_{1:R} = \hat{\mathbf{z}}_{1:R-1} + \hat{\mathbf{z}}_R$ (line 23). Otherwise, $\hat{\mathbf{z}}_{1:R} = \hat{\mathbf{z}}_{1:R-1}$ (line 27). The resulting $\hat{\mathbf{z}}_{1:R}$ is used to update $\hat{\boldsymbol{\beta}}$ in the next iteration.

$\hat{\boldsymbol{\beta}}$ and $\hat{\mathbf{z}}_{1:R}$ are repeatedly updated while reducing the bandwidth according to $h_{R+1} = \delta h_R$ with $0 < \delta < 1$ (line 30) and updating prior variance $\tau_{R+1}^2$, which regularizes the process (line 31).[2] If the validation SSE does not decrease for $Q_{\max}$ consecutive iterations (lines 5–34), the resulting $\hat{\boldsymbol{\beta}}$ and $\hat{\mathbf{z}}_1, \ldots, \hat{\mathbf{z}}_R$ are taken as the final outputs and used in the second HV.

3.1.4. Second HV

Although the first HV optimizes $\hat{\mathbf{z}}_1, \ldots, \hat{\mathbf{z}}_R$ to minimize $SSE_R = \left\| \mathbf{y}_{(v)} - \mathbf{X}_{(v)}\hat{\boldsymbol{\beta}} - \hat{\mathbf{z}}_{1:R(v)} \right\|^2$, the solution would be sub-optimal because they are sequentially optimized, not simultaneously. Therefore, we replace the estimate $\hat{\mathbf{z}}_r$ with an adjusted estimate $\hat{\mathbf{z}}_r^* = \alpha_r \hat{\mathbf{z}}_r$. The adjustment factor is specified as $\alpha_r = \theta_1 exp(-\theta_2 h_r)$, which is designed to reduce bias associated with scale or bandwidth. The parameters $\theta_1$ and $\theta_2$ are optimized via HV minimizing $SSE_R^* = \left\| \mathbf{y}_{(v)} - \mathbf{X}_{(v)}\hat{\boldsymbol{\beta}} - \hat{\mathbf{z}}_{1:R(v)}^* \right\|^2$ (line 34). If $\hat{\theta}_1 = 1$ and $\hat{\theta}_2 \to 0$, then $\alpha_r$ becomes uniformly equal to one, and no adjustment is applied. Otherwise, the second HV further reduces the validation error.

Finally, the predictive value at site $s_0$ is obtained as (line 35)

$$\hat{y}(s_0) = \mathbf{x}(s_0)\hat{\boldsymbol{\beta}} + \hat{z}_{1:R}^*(s_0), \tag{13}$$

where $\hat{z}_{1:R}^*(s_0) = \sum_{r=1}^{R} \hat{\alpha}_r \hat{z}_r(s_0)$. The uncertainty of the prediction is evaluated in Section 3.3.

[2] The variance $\tau_{c_{R+1}}^2$ of the Gaussian prior $\mu_{c_{R+1}} \sim N(0, \tau_{c_{R+1}}^2)$ is given by the sample variance of $\hat{\mu}_{c_R}$ estimated in the previous iteration, under the assumption that the variance explained at resolution $R+1$ is roughly the same as that at resolution $R$. The assumption is acceptable as long as the gap between $h_{R+1}$ and $h_R$ is small.

## 3.2. Non-linear spatial modeling

### 3.2.1. Model

This section considers the following non-linear model:

$$\mathbf{y} = \mathbf{X}\boldsymbol{\beta} + \mathbf{f}(\mathbf{X}_f) + \mathbf{z}_{1:R} + \mathbf{e}, \qquad \mathbf{e} \sim N(\mathbf{0}, \sigma^2\mathbf{I}). \tag{14}$$

This non-linear model contains a vector $\mathbf{f}(\mathbf{X}_f)$ whose elements are determined by a function that captures non-linearity and/or higher-order interactions in the covariate matrix $\mathbf{X}_f$. While the function for $\mathbf{f}(\mathbf{X}_f)$ may be specified using random forests, gradient-boosting trees, or neural networks, we consider the random forest model as an example. We assumed the following model for the prediction site $s_0$.

$$y(s_0) = \mathbf{x}(s_0)\boldsymbol{\beta} + f(\mathbf{x}_f(s_0)) + z_{1:R}(s_0) + e(s_0), \quad e(s_0) \sim N(0, \sigma^2), \tag{15}$$

where $\mathbf{x}_f(s_0)$ is a covariate vector for the non-linear term $f(\mathbf{x}_f(s_0))$.

### 3.2.2. Learning algorithm

Based on the principle of progressive learning that simpler terms should be trained earlier (Section 3.1), the non-linear term $\mathbf{f}(\mathbf{X}_f)$, which accommodates non-smooth complex patterns, should be estimated after $\hat{\mathbf{z}}_1, \dots, \hat{\mathbf{z}}_R$, which describes smooth patterns. The resulting learning procedure is as follows:

(1) Apply the two HVs and obtain the optimal predictor $\hat{\mathbf{y}} = \mathbf{X}\hat{\boldsymbol{\beta}} + \hat{\mathbf{z}}_{1:R}$ as detailed in Section 3.1,

(2) Optimize the estimate $\hat{\mathbf{f}}(\mathbf{X}_f)$ using another HV minimizing the validation SSE of the predictor $\hat{\mathbf{y}}^* = \hat{\mathbf{y}} + \hat{\mathbf{f}}(\mathbf{X}_f)$ using the same training and validation samples.

(3) If the validation SSE is reduced, $\hat{\mathbf{y}}^*$ is the output predictor; otherwise, $\hat{\mathbf{y}}$ is the output predictor.

Because step (2) is a conventional HV, it can be easily implemented using off-the-shelf software packages. This simple algorithm is useful for estimating models that consider linear and non-linear trends as well as latent spatial processes.

Finally, the predictive value at site $s_0$ is $\hat{y}(s_0) = \mathbf{x}(s_0)\hat{\boldsymbol{\beta}} + \hat{f}(\mathbf{x}_f(s_0)) + \hat{z}^*_{1:R}(s_0)$. The next section explains the evaluation of the predictive variance.

## 3.3. Uncertainty modeling

Predictive values can be efficiently resampled by using $\hat{z}_1(s_0), \ldots, \hat{z}_R(s_0)$ and $\hat{\sigma}^2_r(s_0), \ldots, \hat{\sigma}^2_r(s_0)$, which are obtained during the learning procedure described above.

For the linear spatial model (Eq. 11), the *b*-th resampling was generated as follows:

(1) Resample $\boldsymbol{\beta}^{(b)} \sim N(\hat{\boldsymbol{\beta}}, Var[\hat{\boldsymbol{\beta}}])$, where $Var[\hat{\boldsymbol{\beta}}] = \hat{\sigma}^2(\mathbf{X}'\mathbf{X})^{-1}$ and $\hat{\sigma}^2 = \hat{\mathbf{e}}_R{}'\hat{\mathbf{e}}_R/(N-K)$.

(2) Resample $z^{(b)}_r(s_0) \sim N(\hat{z}_r(s_0), \hat{\sigma}^2_r(s_0))$ for each $r$ and evaluate $z^{(b)}_{1:R}(s_0) = \sum_{r=1}^{R} \hat{\alpha}_r z^{(b)}_r(s_0)$.[3]

(3) Resample $e^{(b)}_R(s_0)$ from $\hat{e}_R(s_1), \ldots, \hat{e}_R(s_N)$ with replacement.[4]

[3] The adjustment parameters $\hat{\theta}_1$ and $\hat{\theta}_2$ in $\hat{\alpha}_{\hat{\boldsymbol{\theta}},r}$ are fixed by their estimates since our preliminary analysis suggested that their uncertainty is small and has little impact on the analysis result.

[4] The residuals $\hat{e}_R(s_1), \ldots, \hat{e}_R(s_N)$ are nearly independent and exchangeable since spatial dependence is filtered by the spatial process term $z^*_{1:R}(s_i)$ (Murakami and Griffith, 2015)

(4) Simulate the predictive value as $y^{(b)}(s_0) = \mathbf{x}'(s_0)\boldsymbol{\beta}^{(b)} + z_{1:R}^{(b)}(s_0) + e_R^{(b)}(s_0)$.

In the case of a non-linear spatial model (Eq. 14), step (4) is replaced by steps (5) and (6).

(5) Resample $f(x_f(s_0))^{(b)}$ from the trained model. For example, if $f(x_f(s_0))^{(b)}$ is a random forest model, then quantile regression forests (Meinshausen and Ridgeway, 2006) are available.

(6) Simulate the predictive value as $y^{(b)}(s_0) = \mathbf{x}'(s_0)\boldsymbol{\beta}^{(b)} + f(x_f(s_0))^{(b)} + z_{1:R}^{(b)}(s_0) + e_R^{(b)}(s_0)$.

Replicates $y^{(1)}(s_0), \ldots, y^{(B)}(s_0)$ are obtained by iterating this simulation $B$ times (e.g., $B$ = 200). They are useful for evaluating the predictive variance, interval, and other uncertainty measures at the prediction site $s_0$.

## 3.4. Properties

The CFSM framework differs from existing spatial modeling approaches in several aspects. First, while spatial dependence is typically represented through covariance modeling, CFSM relies on local variance weighting. A related approach is the geographically weighted regression (GWR; Brunsdon et al., 1996; Fotheringham et al., 2017; Lu et al., 2018; Comber et al., 2023) which also relies on local weighting. The local models estimated from GWR are directly interpreted. By contrast, the local models in CFSM are not directly interpreted; instead, they are aggregated to construct a spatial process. In spatial statistics, little attention has been paid to such stochastic processes based on local model aggregation, making CFSM a distinctive contribution.

Second, the CFSM framework does not require explicit matrix inversion, which is a computational bottleneck in conventional spatial models, particularly in GPs. In addition, within each scale, the local models are estimated independently, making this step naturally parallelizable. Thus, CFSM provides a computationally efficient alternative for spatial process modeling.

Third, CFSM optimizes the model using HV, in contrast to the likelihood-based inference commonly employed in spatial statistics. Because many machine learning algorithms are also optimized via validation, the use of HV facilitates much more easy integration with these algorithms than with conventional likelihood-based models.

Based on Eqs. (2) and (9), our predictive mean can be expressed as $\hat{z}_{1:R}(s_i) = \sum_{r=1}^{R} \frac{\alpha_r}{\sum_{c_r=1}^{C_r} w_{c_r}(s_i)} \sum_{c_r=1}^{C_r} w_{c_r}(s_i)\hat{\mu}_{c_r}$, which represents a multiscale sum of kernels. In this sense, our model is a basis function model that uses kernels or other basis functions to represent spatial processes (e.g., Cressie and Johanesson, 2008; Wood, 2017). Although basis function models are computationally efficient and widely used (see Cressie et al., 2022), they suffer from a degeneracy problem, a substantial loss of predictive accuracy when the number of kernels/basis functions $C$ is insufficient (Stein, 2014). Unfortunately, conventional models, which assume a fixed $C$ value that is much smaller than $N$, cannot easily overcome this challenge without loss of computational efficiency. Several studies have reported their limited predictive performance (e.g., Heaton, 2019; Hazra et al., 2025).

In contrast, CFSM optimizes the total number of kernels $\sum_{r=1}^{R} C_r$ rather than pre-specifying it, by sequentially adding $C_r$ kernels at each scale. This sequential optimization naturally avoids the simultaneous estimation of all model components, which makes conventional models slow when $C$ is large.[5] Furthermore, because the optimization is performed through HV, CFSM can incorporate more kernels than $N$ if it improves the validation score. These properties enable CFSM to use a substantially larger number of kernels than conventional basis function models without incurring computational difficulties. Consequently, CFSM provides an effective remedy for the degeneracy problem.

In summary, the proposed framework is a unique, scalable, and machine learning-compatible spatial process. Section 5 investigates predictive performance through Monte Carlo experiments.

## 4. Monte Carlo experiment

Section 4.1 presents a Monte Carlo experiment to compare our linear spatial model with alternative models. Section 4.2 reports a similar experiment for the non-linear spatial model, using random forest as a representative example.

[5] Typically, an inversion of $C \times C$ design matrix whose computational complexity equals $O(C^3)$ is required (see Wood, 2017).

## 4.1. Experiment for the linear spatial model

### 4.1.1. Outline

In this section, the predictive accuracy of the proposed method is evaluated using synthetic samples generated from the following model:

$$
\begin{aligned}
y(s_i) &= \beta_0 + \beta_1 x_1(s_i) + \beta_2 x_2(s_i) + z(s_i) + e(s_i), \qquad e(s_i) \sim N(0,1), \\
z(s_i) &= \sum_{j=1}^{N} w_h^{scale}(d_{ij})\, u(s_j), \qquad u(s_j) \sim N(0,2^2),
\end{aligned}
\tag{16}
$$

where $\{\beta_0, \beta_1, \beta_2\} = \{1, 2, -0.5\}$, and $w_h^{scale}(d_{ij}) = \frac{w_h(d_{ij})}{\sum_j w_h(d_{ij})}$ is a row-standardized weight. Here, $w_h(d_{ij}) = \exp(-d_{ij}^2/h^2)$ is the Gaussian kernel, and $d_{ij}$ is the Euclidean distance between sites $s_i$ and $s_j$. The spatial coordinates of data $y(s_i)$ were randomly selected within the region [0, 10] × [0, 10]. Because spatially correlation is common in many real-world cases, the covariables were also modeled as spatial processes with moderate scale, using a bandwidth value of 1.0, defined as follows:

$$
\begin{aligned}
x_k(s_i) &= 0.5 z_k(s_i) + 0.5 e_k(s_i), \qquad e_k(s_i) \sim N(0,1), \\
z_k(s_i) &= \sum_{j=1}^{N} w_1^{scale}(d_{ij})\, u_k(s_j), \qquad u_k(s_j) \sim N(0,1).
\end{aligned}
\tag{17}
$$

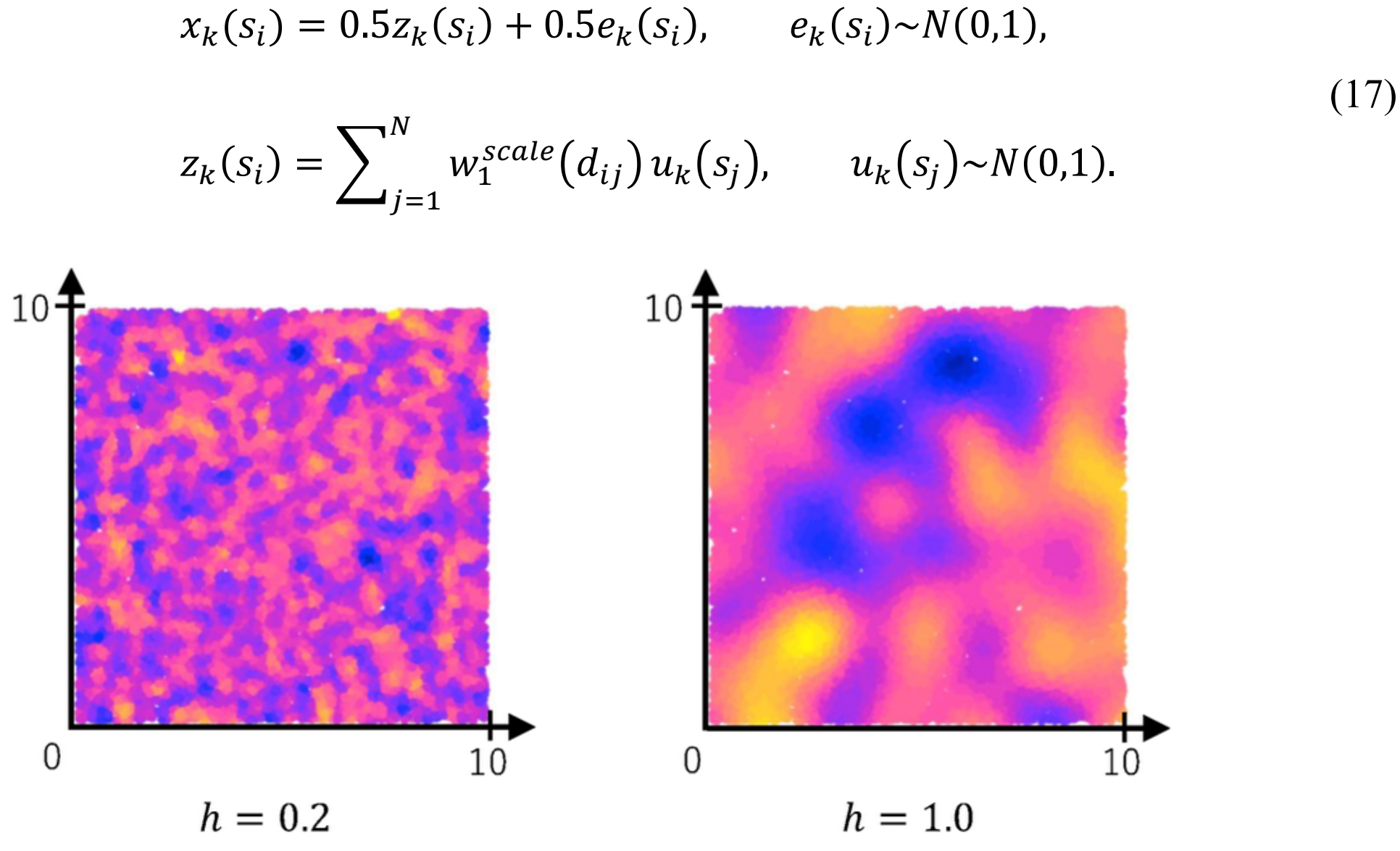

Figure 7. Spatial process $z(s_i)$ simulated with $h \in \{0.2, 1.0\}$.

The synthetic data is generated 200 times for each case with different training sample sizes $N \in \{500,1000,2000,3000,6000,12000,20000\}$ and bandwidths $h \in \{0.2, 1.0\}$. As illustrated in Figure 7, the spatial process $z(s_i)$ exhibited small-scale variations when $h = 0.2$ and large-scale variations when $h = 1.0$. These two cases are useful for examining how accurately an algorithm estimates the scale of a process.

In each trial, 1,000 additional test samples were generated in the study region and used to evaluate the predictive root mean squared error (RMSE) and mean absolute error (MAE).

$$RMSE = \sqrt{\frac{1}{N}\sum_{i=1}^{N}\left(y(s_i) - \hat{y}(s_i)\right)^2}, \tag{18}$$

$$MAE = \frac{1}{N}\sum_{i=1}^{N}|y(s_i) - \hat{y}(s_i)|. \tag{19}$$

RMSE and MAE measure the accuracy of the predictive value. RMSE is more sensitive to outliers owing to squaring, whereas MAE treats all errors equally. The RMSE and MAE values of our linear spatial model optimized by the CFSM are compared with those of several alternative methods, including:

- The conventional spatial GP (GP) with the Gaussian kernel (with nugget effects) for modeling spatially decaying covariance
- A generalized additive model (GAM), which assumes a low rank spatial GP for residual spatial modeling (Kammann and Wand, 2003),

- The conjugate nearest neighbor GP (NNGP) with Gaussian covariance kernel, whose variance and bandwidth parameters are optimized by five-fold cross-validation (Finley et al., 2019), and
- The stochastic partial differential equation (SPDE) method[6] (see Section 1).

The GP was optimized through non-linear weighted least squares estimation (Cressie, 2015), which was implemented in the R package gstat (https://cran.r-project.org/web/packages/gstat/index.html). Considering computational complexity, we apply GP only in cases with $N \leq 3000$. The GAM was implemented using the R package mgcv (https://cran.r-project.org/web/packages/mgcv/index.html), and the number of basis functions followed the default settings in the package. NNGP and SPDE were implemented using the spNNGP (https://cran.r-project.org/web/packages/spNNGP/index.html) and INLA (https://www.r-inla.org/home) packages, respectively. All these methods are implemented without parallelization.

#### 4.1.2. Result

Figure 8 shows the medians of the RMSE and MAE values. When the true process had a large-scale pattern with $h = 1.0$, GP performed best, as expected. For $N \leq 3000$, the SPDE and CFSM methods produced comparable accuracy, while for $3000 < N$, they achieved the smallest

[6] The maximum edge length of the spatial graph used to approximate GP was set to 0.25, which is considered sufficiently small relative to the size of the study region [0, 10] × [0, 10].

RMSE and MAE values. The accuracy of the proposed method was confirmed for large-scale spatial modeling.

In the case of a small-scale process with $h = 0.2$, GP failed to capture the local patterns, and the resulting process was less accurate. This may be due to the difficulty in distinguishing small-scale variations in $z(s_i)$ and $e(s_i)$. The NNGP accurately predicted the small-scale patterns. Because NNGP is a nearest-neighbor-based approximation, this result is reasonable. The CFSM framework yielded even smaller errors owing to the sequential learning procedure that optimized the scale. As shown in Figure 8, only CFSM achieved accurate predictions for both large-scale and small-scale processes.

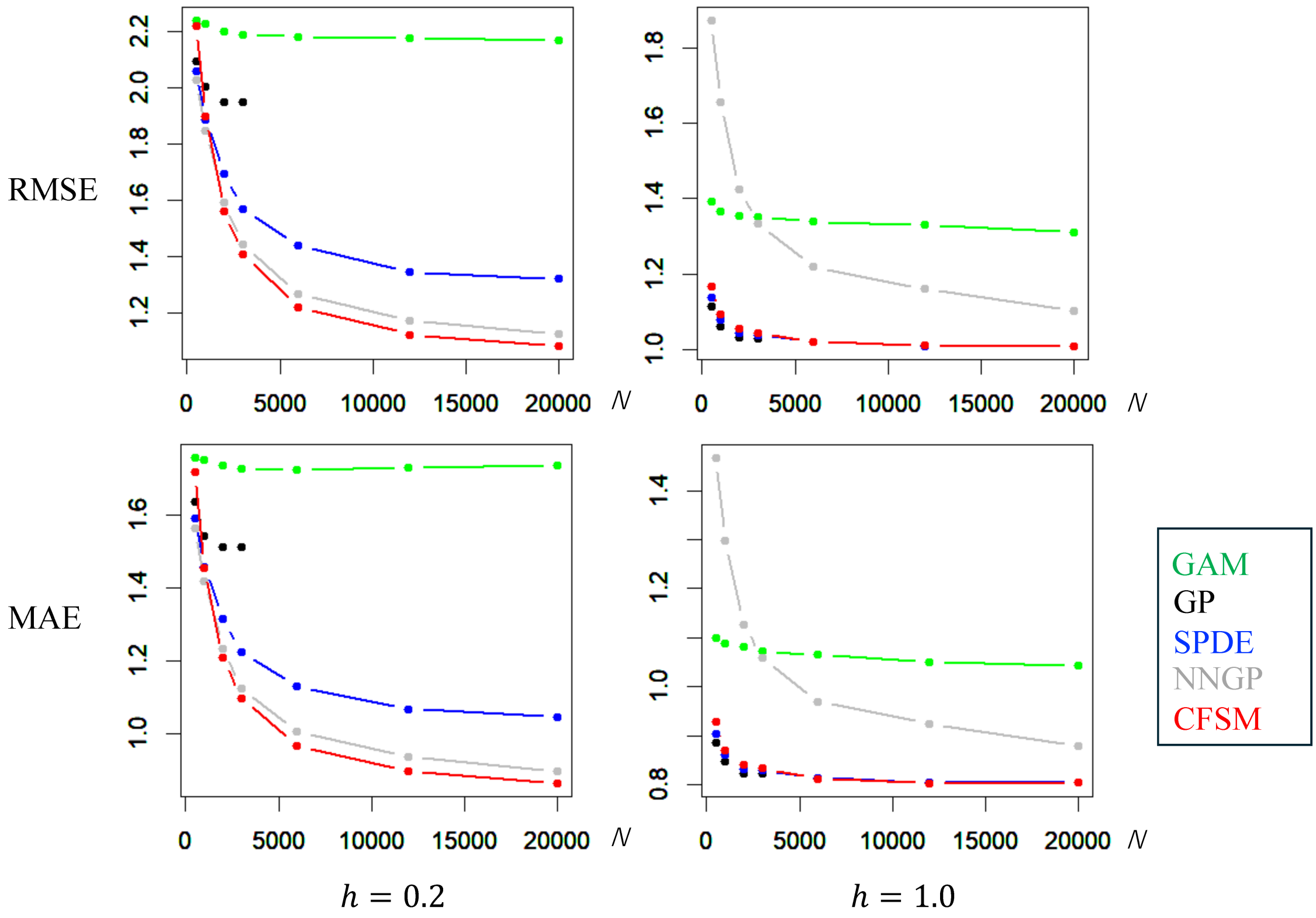

Figure 8. Medians of the RMSE and MAE values (True: linear spatial model).

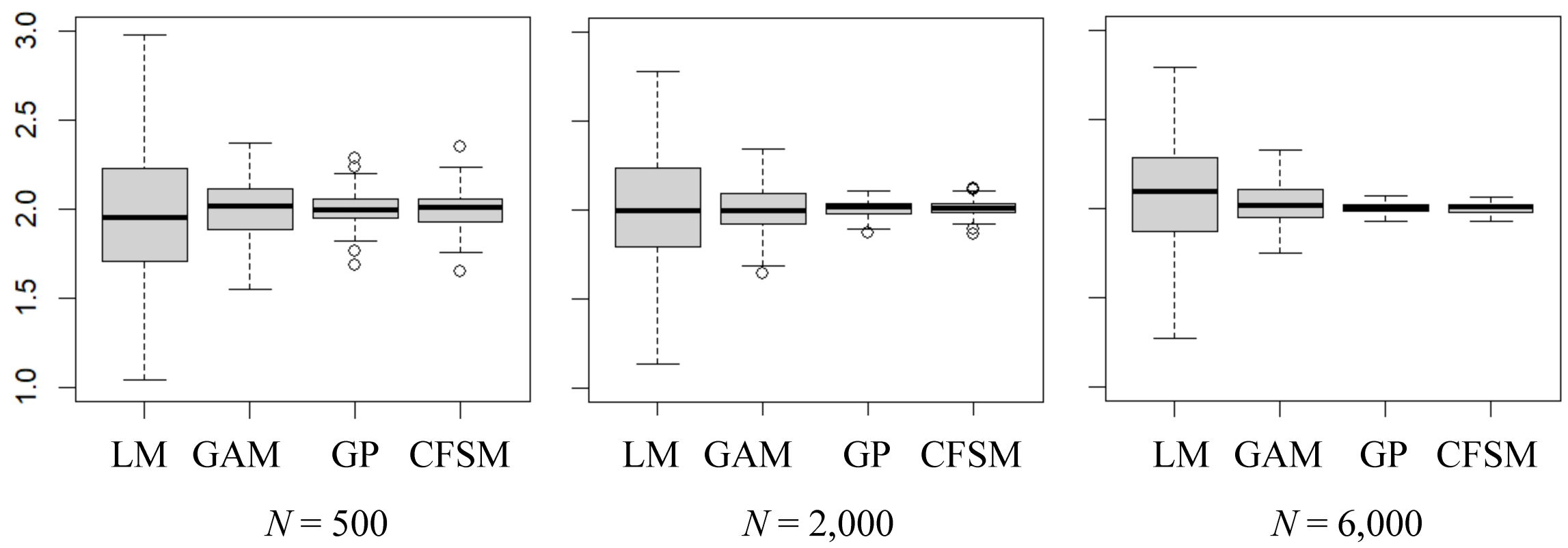

Figure 9. Boxplots of the estimated $\hat{\beta}_1$ values (true: $\beta_1 = 2.0$) for $N = 500$, 2,000, and 6,000.

Figure 9 compares the boxplots of the estimated values of the regression coefficient $\beta_1$. Results for $\beta_0$ and $\beta_2$, which were similar to those for $\beta_1$, are omitted for simplicity. Here, the CFSM framework was compared with the GAM and GP, which are widely used for regression analysis, as well as with basic linear regression (LM). Owing to the ignorance of spatial dependence, which results in the loss of statistical efficiency (LeSage and Pace, 2009), the coefficient estimates of LM varied considerably across trials and were less accurate. The coefficient estimates of the GAM were also less accurate because of the low predictive model accuracy reported in Figure 9. By contrast, the estimates of CFSM consistently remained close to the true value of $\beta_1 = 2.0$, similar to GP. This result highlights the usefulness of CFSM for accurate coefficient estimation.

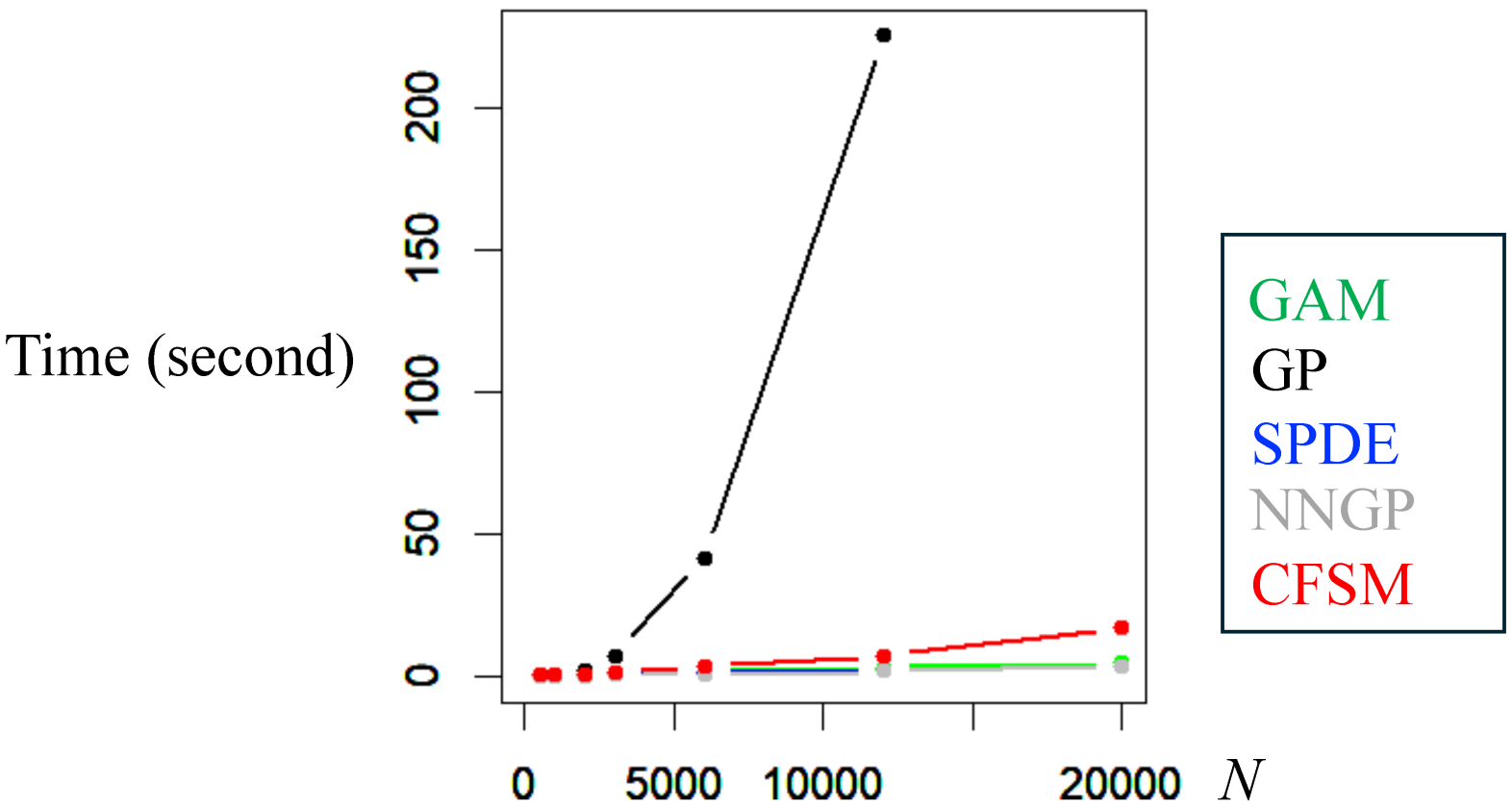

Figure 10. Median computation time. GP was not estimated when $N = 20{,}000$ due to memory limit.

Figure 10 shows a comparison of the median computation times over the five iterations. All computations were performed using R 4.5.1, on a Windows 11 Pro computer with a 13th Gen Intel Core i7-13700 CPU and 16 GB of memory. As expected, the computation time of GP rapidly increased with *N*. In contrast, the computation time of the CFSM was much shorter and increased only linearly with respect to *N*, confirming its computational efficiency. For example, for $N$ = 12,000, GP took 225.7 s on average, while CFSM required only 7.4 seconds. Although it is slower than SPDE, NNGP, and GAM, it remains useful for large samples given its ease of parallelization.

The spatial processes $z(s_i), z_1(s_i), \dots, z_K(s_i)$ in the synthetic data were generated using a moving average process, which allowed the generation of large samples. For reference, Appendix 4 summarizes the results obtained when these spatial processes are replaced with GPs for cases with $N \leq 6{,}000$ where computation is feasible. The overall results are similar to those reported previously.

In cases with large-scale processes ($h = 1.0$), GP performed best (as expected), whereas CFSM showed comparable accuracy, indicating smaller predictive errors than the other approximate GPs. In the case of small-scale processes ($h = 0.2$), the accuracy was comparable to that of NNGP, which yielded the best performance. Notably, CFSM was the only method that consistently achieved an accuracy comparable to that of the best-performing method in both scenarios.

## 4.2. Experiment for the non-linear spatial model

### 4.2.1. Outline

To evaluate the performance of the non-linear spatial model, we generated simulation samples according to

$$y(s_i) = \beta_0 + \beta_1 \exp(x_1(s_i)) + \beta_2 x_2^+(s_i) + z(s_i) + e(s_i), \qquad e(s_i) \sim N(0,1), \tag{20}$$

where $x_2^+(s_i) = \max(x_2(s_i), 0)$. The covariates $x_1(s_i), x_2(s_i)$ and spatial process $z(s_i)$ were generated in the same manner as described in a previous section. We set $\beta_0 = 1$, and choose $\beta_1$ and $\beta_2$ such that $Var[\beta_1 \exp(x_1(s_i))] = Var[\beta_2 x_2^+(s_i)] = 2^2$, which equals $Var[z(s_i)]$. Because of the non-linear effects of $x_1(s_i)$ and $x_2(s_i)$, the non-linear CFSM framework is appropriate in this case.

Samples were generated 200 times for each case with $N \in \{500,1000,2000,3000,6000,12000,20000\}$ and kernel bandwidth $h = 1.0$. In each case, the RMSE and MAE of the following models were compared: our linear spatial model (CFSM), our non-linear spatial model with additional training by random forest (CFSM-RF), random forest (RF) based on five-fold cross-validation, implemented in the ranger (https://cran.r-project.org/web/packages/ranger/index.html) R package, and the SPDE approach, which was found to be accurate in the case of $h = 1.0$ in a previous section. In the RF and RF training parts of CFSM-RF, spatial coordinates were considered as additional covariates to capture spatial patterns. Among the RF hyperparameters, the number of candidate predictors at each split and minimum number of samples per terminal node can significantly influence the analysis results. They were optimized through a grid search, while the number of trees was fixed at 500, as the results were found to be similar when the number was increased.

### 4.2.2. Predictive accuracy

Figure 11 shows a comparison of predictive accuracy. As expected, the SPDE and CFSM, which are linear models, failed to capture non-linearity and thus showed lower accuracy. By contrast, RF, which can model non-linearity, tended to be more accurate. Our CFSM-RF achieved the lowest RMSE and MAE values across all cases, demonstrating its superior predictive accuracy.

Table 1 compares the accuracy of the four models in predicting the data generated by the linear model (Eqs. 15–16) with $h = 1.0$. RF performed poorly because it failed to exploit the underlying linearity. In contrast, CFSM-RF produced the same results as the CFSM, confirming its accuracy. These results confirm the predictive accuracy of CFSM-RF for the linear and non-linear cases.

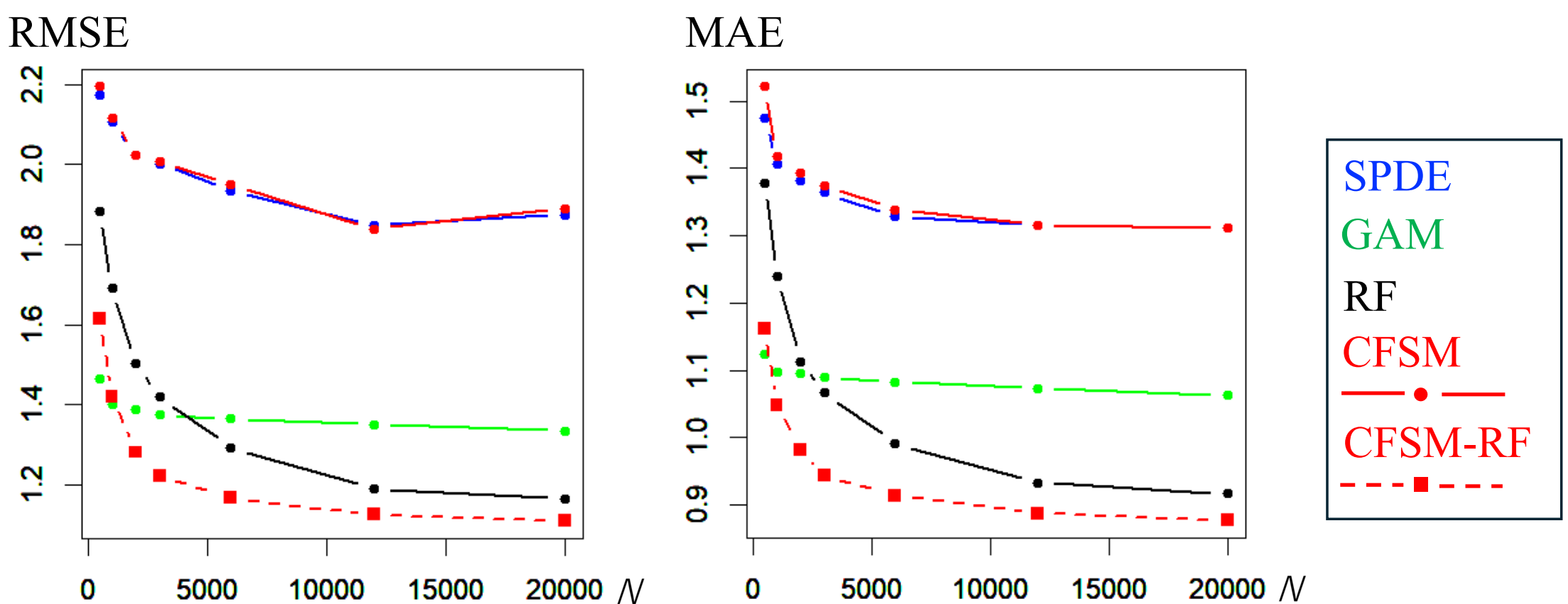

Figure 11. Medians of the RMSE (left) and MAE (right) values (True: non-linear spatial model).

Table 1. Medians of the RMSE and MAE values (True: linear spatial model with $N$ =6,000).

| | SPDE | GAM | RF | CFSM | CFSM-RF |
|---|---|---|---|---|---|
| RMSE | 1.020 | 1.331 | 1.184 | 1.020 | 1.020 |
| MAE | 0.814 | 1.064 | 0.941 | 0.812 | 0.812 |

### 4.2.3. Partial dependence of the covariates

Figure 12 presents accumulated local effects (ALE) plots showing the average effects of $x_1(s_i)$ and $x_2(s_i)$ on the model predictions (Apley and Zhu, 2020). As summarized in Table 2, CFSM-RF estimated the non-linear effect $f(x_1(s_i)) = \beta_1 \exp(x_1(s_i))$ more accurately than RF, likely because the pre-trained linear trend helps capture the rapid increase in $f(x_1(s_i))$ for large $x_1(s_i)$ values (Figure 12 (left)). In contrast, CFSM-RF showed lower accuracy for $f(x_2(s_i)) = \beta_2 x_2^+(s_i)$ as the pre-trained negative linear trend leads to underestimation for large $x_2(s_i)$ values (Figure 12 (right)). Thus, near the edges of the ALE plot, such pre-trained linear trends can distort non-linear patterns owing to limited samples, making it difficult to capture the true (non-linear) trend. Nevertheless, the overall estimation results of CFSM-RF are close to those of the conventional RF and remain interpretable.

To mitigate the edge effects in the ALE plots, the plots may be interpreted only within the 95-th percentile interval (Figure 12). A promising direction for future work is to extend the linear pre-training step in CFSM-RF to better capture non-linear trends, for example, by incorporating spline-based models.

In summary, the linear CFSM and non-linear CFSM (CFSM-RF) provide highly accurate prediction and coefficient estimates. Furthermore, the CFSM-RF model also provided interpretable non-linear effects. The next section illustrates their practical utility through an empirical application.

Table 2. Medians of the RMSE and MAE values (True: linear spatial model with $N$ =6,000).

| | $f(x_1(s_i))$ | | $f(x_2(s_i))$ | |
|---|---|---|---|---|
| | RF | CFSM-RF | RF | CFSM-RF |
| RMSE | 2.452 | 2.181 | 0.347 | 0.376 |
| MAE | 0.370 | 0.363 | 0.184 | 0.232 |

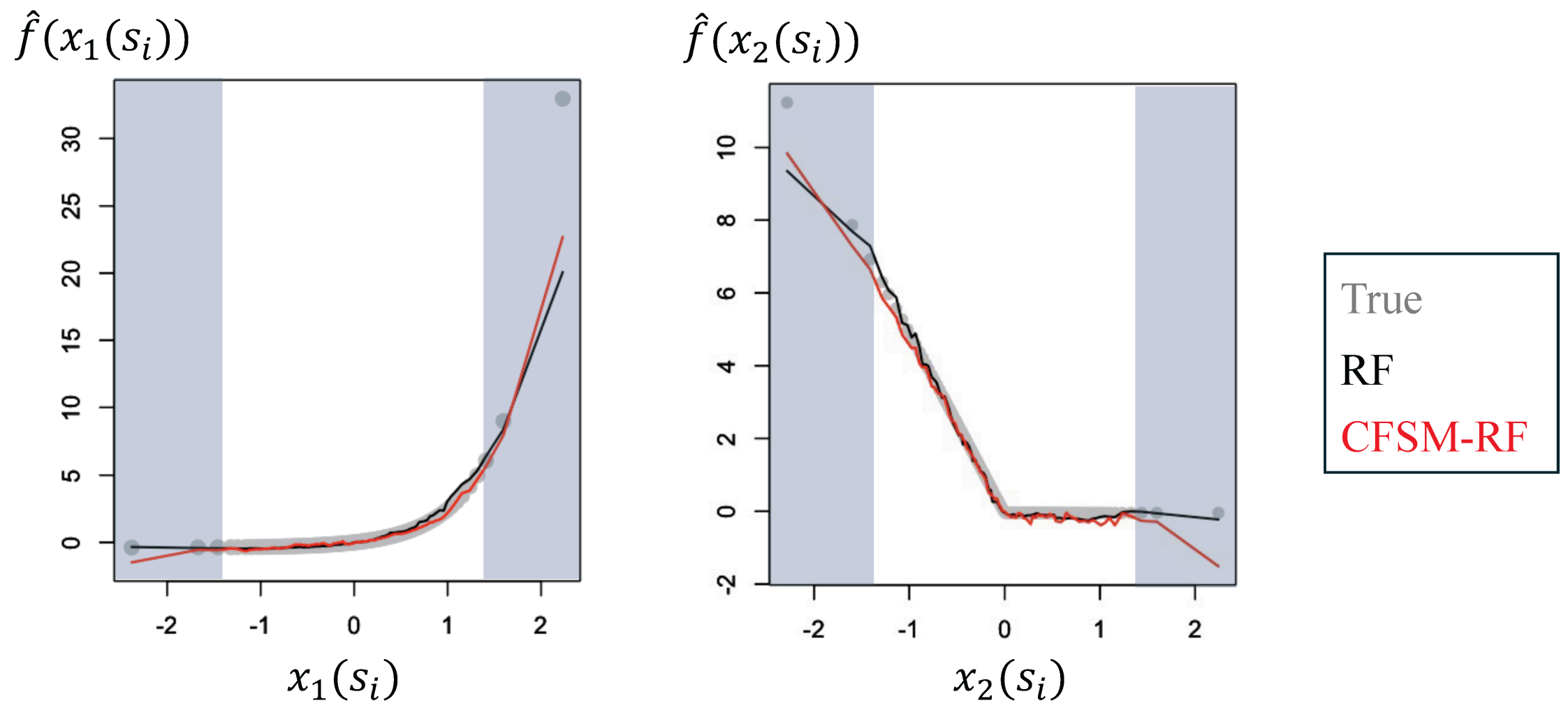

Figure 12. Estimated non-linear effects of $x_1(s_i)$ and $x_2(s_i)$ (*N*=6,000). Regions outside the 95-th percentile interval are shown in blue.

# 5. Application

## 5.1. Outline

This section applies the proposed method to an analysis of residential land prices in 2007 in the Tokyo metropolitan area. The explained variable is the log-transformed residential land price (*N* = 7,497; JPY/m$^2$). As shown in Figure 13, land prices tended to be high in central Tokyo. The study area has a well-developed railway network, and many residents commute to the center by train. As a result, high-price areas stretch along train lines.

The models compared included basic linear regression, RF, GP based on Gaussian and exponential kernels ($GP_{Gau}$, $GP_{Exp}$), linear CFSM based on these kernels ($CFSM_{Gau}$, $CFSM_{Exp}$), and non-linear CFSM additionally trained by RF ($CFSM\text{-}RF_{Gau}$, $CFSM\text{-}RF_{Exp}$).

Their covariables were the Euclidean distance from the nearest railway station (StaDist; km), railway network distance from the nearest station to Tokyo station (TokyoDist; km), and the proportions of agricultural land, forest, wasteland, and river in 500-m grids including the sample site (Agri, Forest, Waste, and River). In addition, for the RF, longitude and latitude were considered as covariables for learning spatial patterns. Yoshida et al. (2024) showed that this RF using spatial coordinates as covariates achieves the highest predictive accuracy among a non-spatial RF and various spatial RF specifications, including nearest neighbors, autoregressive modeling, and spatial basis function-based approaches. Accordingly, the RF adopted in this study serves as an reasonable benchmark for evaluating the predictive accuracy of our method. Data on land prices and covariates are available in the National Land Numerical Information download site (https://nlftp.mlit.go.jp/ksj/).

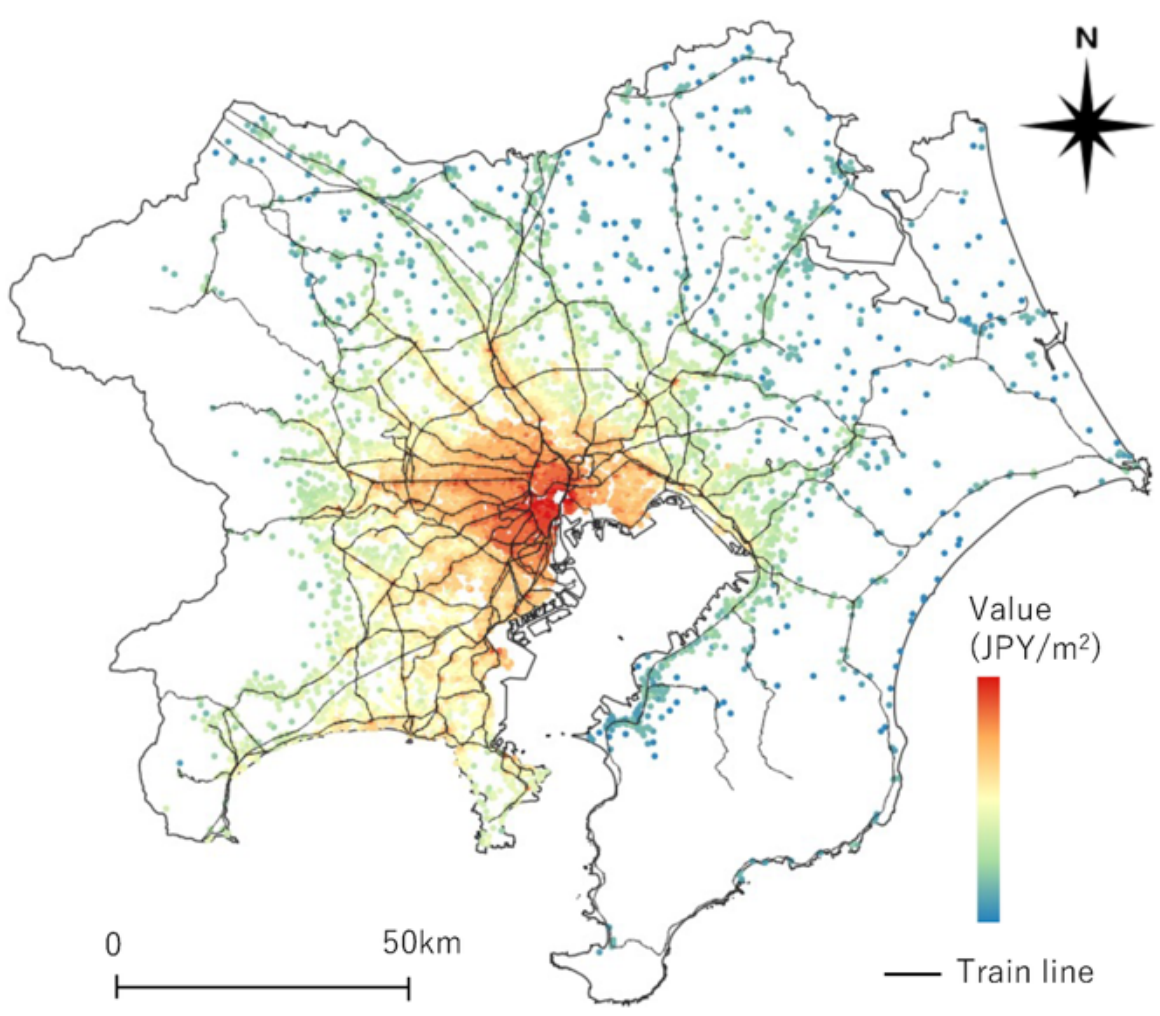

Figure 13. Residential land price in the Tokyo metropolitan area. Central Tokyo, shown at the center of this map, records the highest land prices.

These models are used to predict the land price in each district, called *Cho-Cho-Aza*, whose spatial coordinates are given by their geometric centers.

As illustrated in Figure 1, our multiscale process comprises $R$ single-scale processes. Section 5.2 presents the estimated value of $R$, the bandwidths of the single-scale processes, and other parameters. Then, Section 5.3 reports the prediction results based on the multiscale process, along with its scale-wise decomposition.

## 5.2. Parameter estimation results

Table 3 summarizes the estimated parameters of the linear models. All models indicate statistically significant negative effects of TokyoDist and StationDist, reflecting higher land prices near Tokyo Station and other train stations. The coefficients for Agri, Forest, Waste, and River are also negative, suggesting that non-urban land uses are less attractive than urban land uses that provide urban facilities. These results were consistent.

The coefficients estimated by CFSM were similar to those obtained by GP, with substantially smaller absolute values for Agri, Waste, and River compared to those estimated by the basic linear regression. This implies that our method offers a reasonable alternative to GP for reducing residual spatial dependence and improving the accuracy of coefficient estimation. Furthermore, the standard errors of the CFSM coefficients are generally smaller than those of the GP, which is consistent with its improved accuracy, as indicated by the lower residual standard error.

Notably, the standard deviation of the estimated spatial process in CFSM is comparable to that in the GP, indicating that CFSM yields higher accuracy without incurring variance inflation or overfitting. This improvement stems from its flexibility in modeling multi-scale processes, in contrast to GP, which assumes a single scale with an estimated bandwidth of 48.98 km. The optimal number of scales in CFSM is 40, with the bandwidths ranging from 77.03 km to 1.27 km, indicating the multi-scale structure underlying the land price distribution.

Table 3. Parameter estimates (linear models). SD stands for standard deviation. Spatial SD represents the standard deviation of the spatial process. Regarding GP, the estimated noise variance (i.e., nugget) is displayed in the "Residual SD" row.

| | Linear regression | | GP | | CFSM | |
|---|---|---|---|---|---|---|
| Coefficients | Estimate | Standard error | Estimate | Standard error | Estimate | Standard error |
| Const | 13.317 | $1.13\times10^{-2}$ *** | 13.20 | $2.81\times10^{-1}$ *** | 13.30 | $3.84\times10^{-3}$ *** |
| TokyoDist | -0.026 | $3.12\times10^{-4}$ *** | -0.019 | $1.82\times10^{-2}$ *** | -0.028 | $1.10\times10^{-4}$ *** |
| StationDist | -0.096 | $3.79\times10^{-3}$ *** | -0.136 | $4.52\times10^{-3}$ *** | -0.129 | $1.14\times10^{-3}$ *** |
| Agri | -1.455 | $3.80\times10^{-2}$ *** | -0.697 | $2.84\times10^{-2}$ *** | -0.603 | $1.33\times10^{-2}$ *** |
| Forest | -0.557 | $5.50\times10^{-2}$ *** | -0.485 | $4.19\times10^{-2}$ *** | -0.432 | $1.88\times10^{-2}$ *** |
| Waste | -2.287 | $2.66\times10^{-1}$ *** | -0.283 | $1.74\times10^{-1}$ * | -0.214 | $9.15\times10^{-2}$ ** |
| River | -0.767 | $6.96\times10^{-2}$ *** | -0.345 | $4.56\times10^{-1}$ *** | -0.316 | $2.35\times10^{-2}$ *** |
| Residual SD | 0.471 | | 0.257 | | 0.156 | |
| Spatial SD | - | | 0.504 | | 0.497 | |
| Bandwidth | - | | 48.98 | | 1.27–77.03 | |
| Number of scales | - | | - | | 40 | |

[1)] *, **, and *** indicate statistical significance of 10 %, 5 %, and 1 %, respectively.

The estimated 40 scale-wise processes are summed to form large-scale ($b_r \geq 30$), medium-scale ($30 > b_r \geq 10$), and small-scale ($10 > b_r$) components, as shown in Figure 14. Large-scale processes exhibit high values in the southwestern region. Historically, this region has benefited from the early expansion of railway networks, which has improved accessibility and accelerated sub-urbanization. With the inflow of affluent households and development of high-quality residential environments, including educational and cultural facilities, southwestern area has gradually emerged as a prestigious residential zone. This large-scale pattern likely reflects this tendency. The medium-scale processes showed high values in the outer mountainous areas, possibly reflecting a premium for natural amenities not represented by the covariates. Around central Tokyo, both medium- and small-scale processes exhibited high values, suggesting that Tokyo influences its neighboring areas at multiple spatial scales. In contrast, many other cities show peaks only for small-scale processes. These include prefectural capitals, such as Yokohama, Saitama, and Chiba; high-end residential areas, such as Kichijoji, Kunitachi, and Jiyugaoka; and suburban central cities, such as Kashiwa and Kawagoe (Figure 14). This indicates that the attractiveness of these cities is not captured by the covariates.

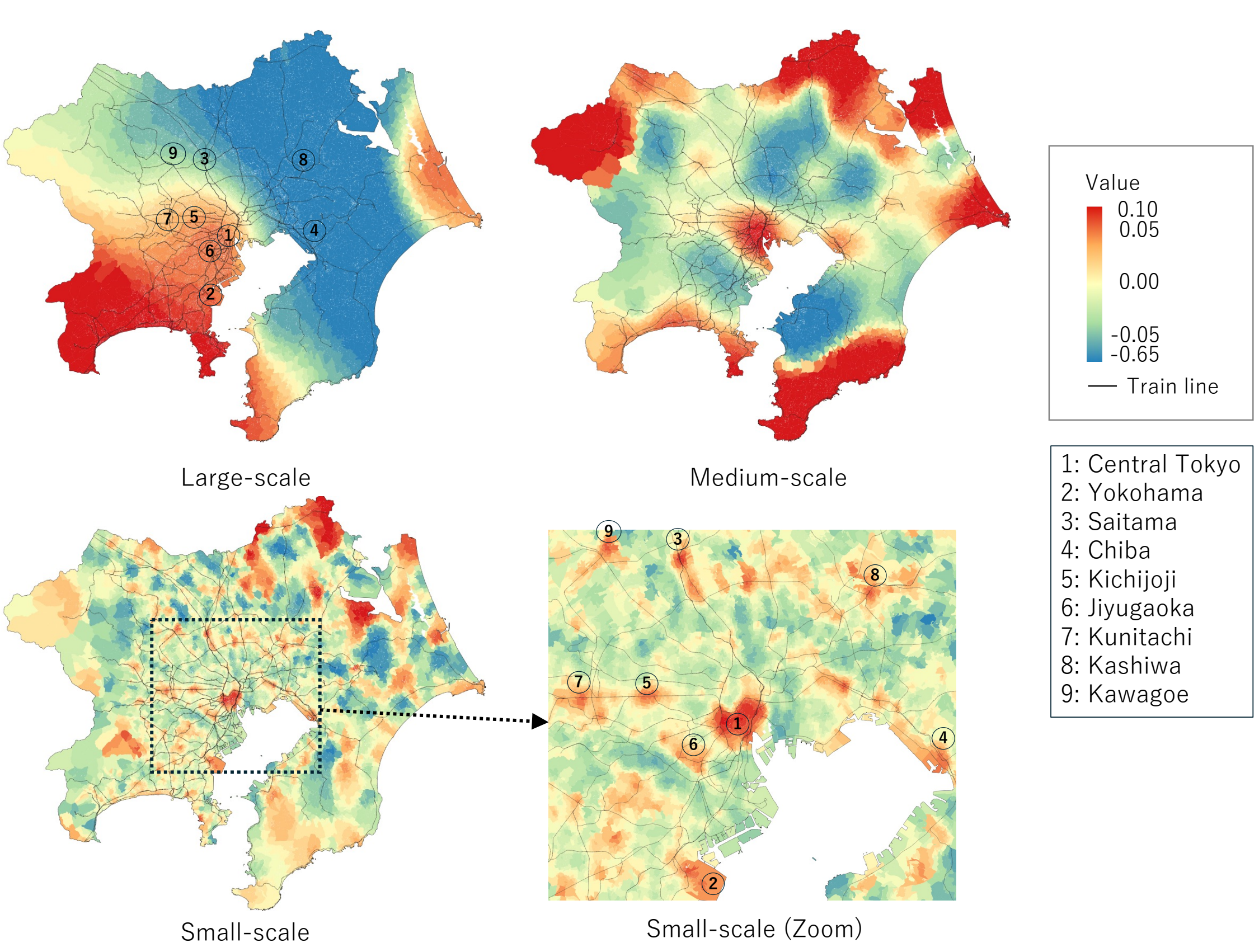

Figure 14. Large-, medium-, and small-scale processes.

For RF and CFSM-RF, Figure 15 presents the ALEs that measure the partial dependence of each covariate on the logged land prices. Both RF and CFSM-RF predicted a rapid decline in logged land prices near railway stations, with a slower decline further away, indicating a strong preference for areas near stations. For TokyoDist, CFSM-RF indicated a linear decline on the log scale, whereas RF suggested a non-linear decay, implying relatively higher prices in areas far from Tokyo Station, including the outer mountainous regions. In CFSM-RF, higher mountainous areas are absorbed by medium-scale residual spatial process, which can be interpreted as premium areas owing to the presence of natural resources (Figure 14). Overall, the effects estimated by CFSM-RF tended to be weaker than those estimated by RF, likely because CFSM-RF explicitly models residual spatial dependence, thereby reducing omitted variable bias.

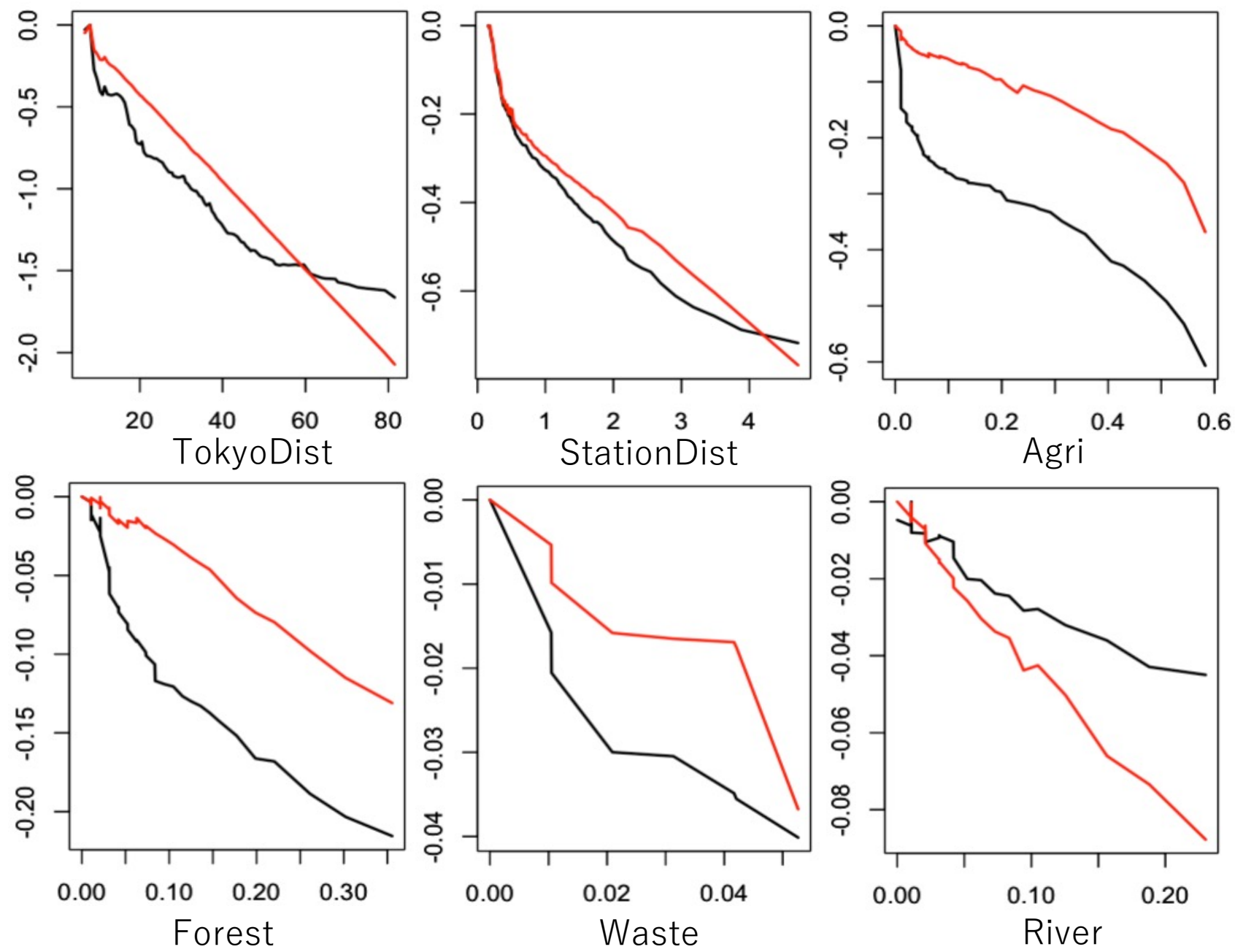

Figure 15. ALE plots (black: RF; red: CFSM-RF). The x-axis shows the covariate value, and the y-axis shows the accumulated local effect. Based on Section 5.2.3, only ALEs within the 95-th percentile interval are displayed.

## 5.3. Spatial prediction result

A 10-fold cross validation was performed to evaluate the RMSE and MAE of the predicted values. In addition, the continuously ranked probability score (CRPS; Gneiting et al., 2005) was evaluated.

$$CRPS = \int_{-\infty}^{\infty} [F(v) - I(y(s_i) \leq v)]\, dv, \tag{21}$$

where $I(y(s_i) \leq v)$ is 1 if $y(s_i) \leq v$, and 0 otherwise. CRPS measures the accuracy of the predictive distribution using the cumulative distribution function $F(\cdot)$ of the predictive distribution.

Table 4. Cross-validation results (Bold: best and second-best models).

| Model | RMSE | MAE | CRPS |
|---|---|---|---|
| Liner regression | 40.889 | 0.356 | 0.349 |
| RF | 19.080 | 0.147 | 0.109 |
| $GP_{Gau}$ | 20.910 | 0.169 | 0.126 |
| $CFSM_{Gau}$ | 19.345 | 0.146 | 0.112 |
| CFSM-$RF_{Gau}$ | **18.513** | **0.141** | **0.108** |
| $GP_{Exp}$ | 19.098 | 0.149 | 0.116 |
| $CFSM_{Exp}$ | 19.063 | 0.144 | 0.110 |
| CFSM-$RF_{Exp}$ | **18.200** | **0.137** | **0.104** |

Table 4 summarizes the cross-validation results. The results indicate that the CFSM and CFSM-RF models outperform the linear regression, RF, and GPs models in terms of RMSE and MAE with high predictive mean accuracies. The gap between CFSM(-RF)$_{Gau}$ and CFSM(-RF)$_{Exp}$ was smaller than that between $GP_{Gau}$ and $GP_{Exp}$, suggesting that the proposed methods are robust to the choice of the kernel function. CFSM(-RF) had small CRPS values, suggesting that the method can accurately model predictive distributions (i.e., uncertainty of the prediction). In addition, CFSM-RF was found to have smaller errors than CFSM for both kernels, indicating that additional learning improved the predictive accuracy.

Here, we focused on exponential-kernel-based specifications based on the comparison results. Figure 16 maps the predicted land prices. The results of $GP_{Exp}$, CFSM $_{Exp}$, and CFSM-$RF_{Exp}$ are quite similar. Compared to the basic linear regression, they indicate higher prices in central Tokyo and its western area, which is a popular residential district due to a number of factors including lower disaster risk, better urban infrastructure, and stronger brand image (Kanno and Shiohama, 2022). In other words, the three models effectively capture the localized price increases in high-end neighborhoods through residual spatial dependence modeling. Compared to RF, they predicted lower land prices in outer mountainous areas whose land price should be lower than that in the other regions. This might be because RF, which ignores spatial dependence, has difficulty in capturing local features in these areas with limited samples. In summary, the results of $GP_{Exp}$, $CFSM_{Exp}$, and CFSM-$RF_{Exp}$ are more reasonable than those of linear regression and RF, highlighting the importance of modeling spatial processes to predict land prices.

Figure 17 compares the predictive standard deviations, which quantify the uncertainty in the predicted values. For the RF, standard deviations were evaluated following Meinshausen and Ridgeway (2006). The basic linear regression indicated consistently high standard deviations, indicating that the low accuracy due to spatial dependence was ignored. RF exhibits spatially discontinuous patterns for the same reason. In contrast, $GP_{Exp}$ produces the smoothest map pattern through spatial process modeling. The predictive standard deviations of $CFSM_{Exp}$, are also smooth but exhibit relatively complex patterns, as the model assumes a complex spatial process composed of $R$ = 40 scale-wise components. This complexity is more pronounced for CFSM-$RF_{Exp}$, which additionally models non-smooth patterns.

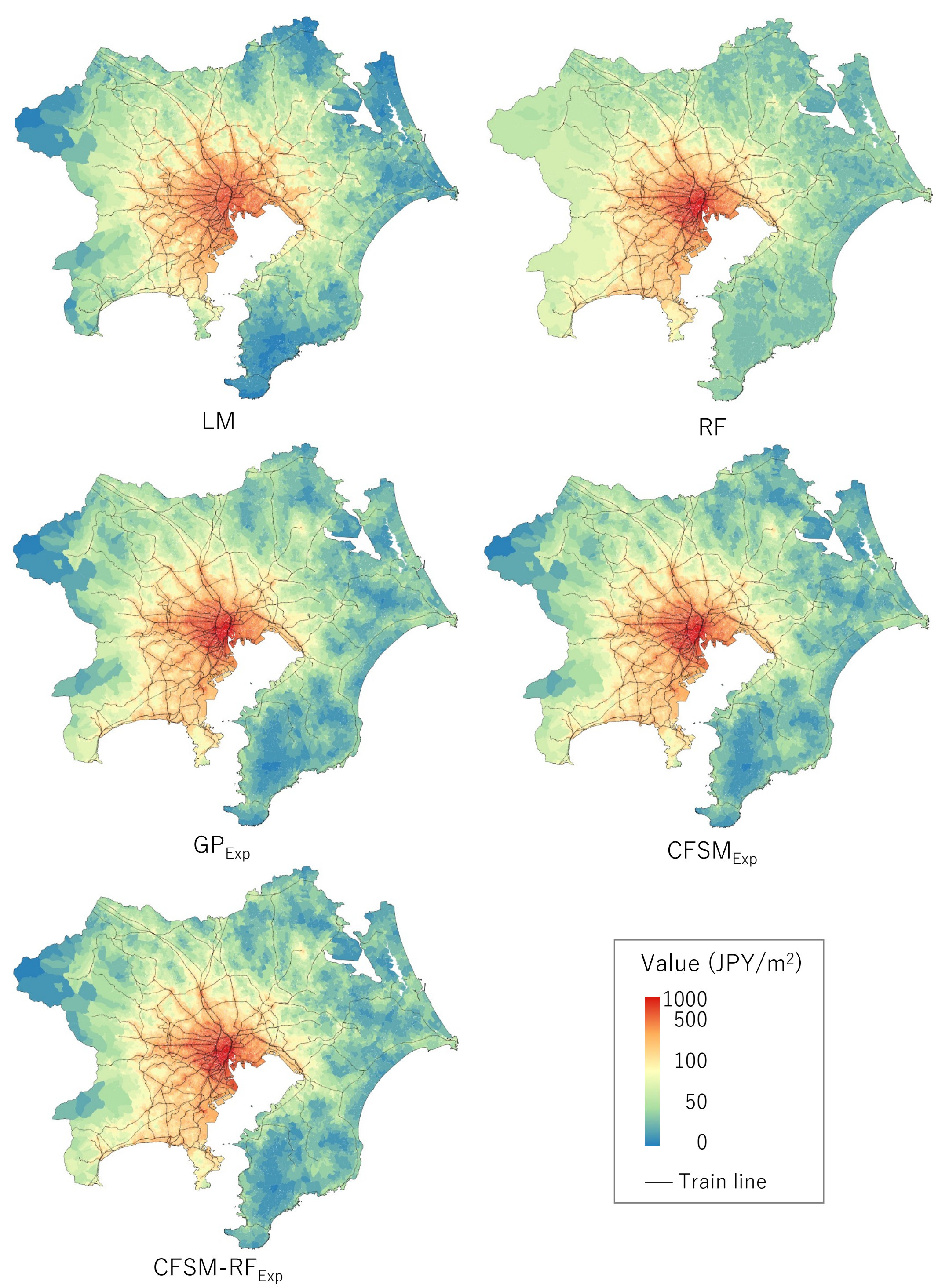

Figure 16. Predicted land prices. Results of $GP_{Gau}$, $CFSM_{Gau}$, and $CFSM\text{-}RF_{Gau}$ are omitted for simplicity.

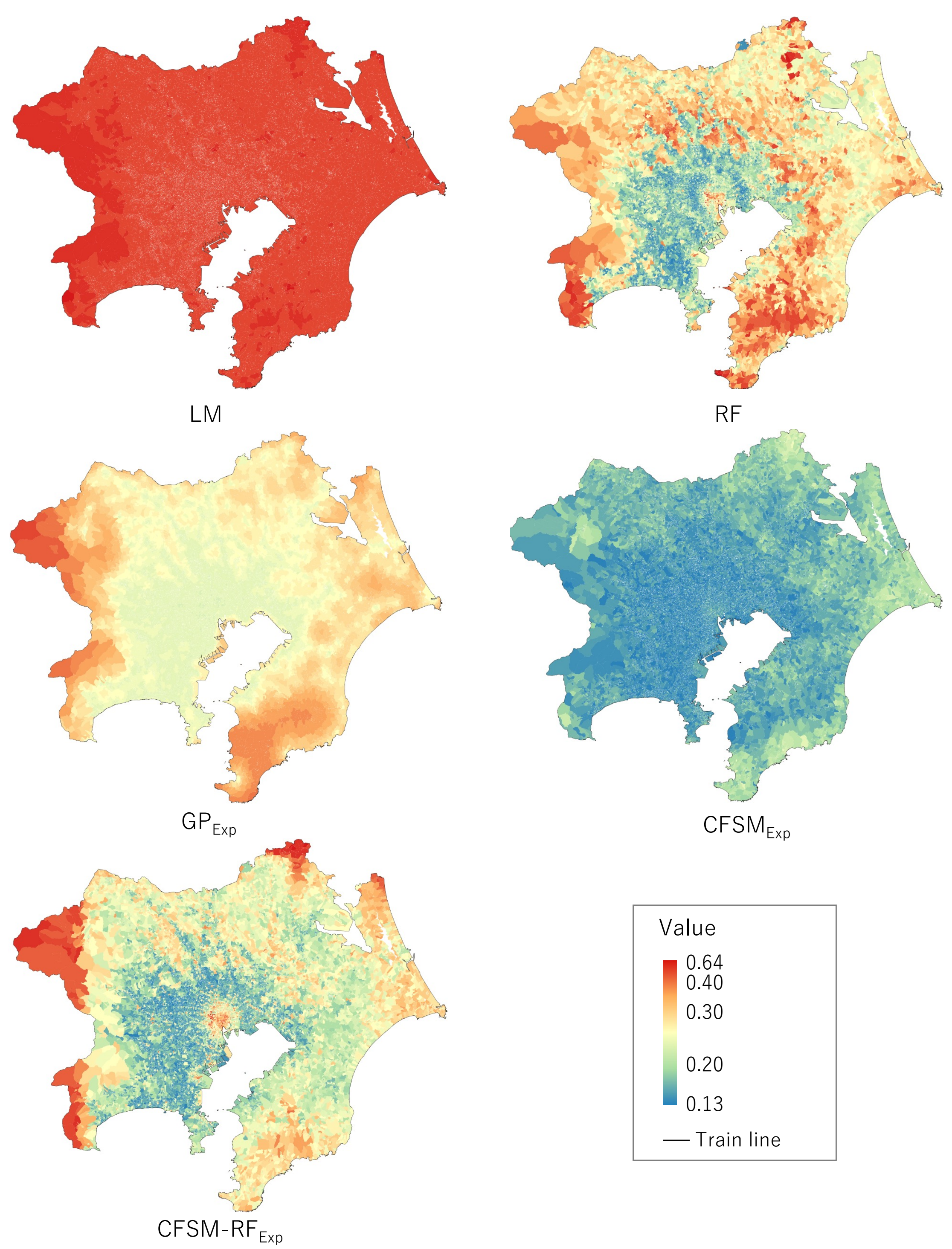

Figure 17. Predictive standard deviations.

Overall, as expected, the predictive standard deviations tended to be large in peripheral areas with sparse sample sites. In addition, the standard deviations of $CFSM_{Exp}$ and CFSM-$RF_{Exp}$ increased in central Tokyo, which is reasonable given the complex urban structure of this area, whereas those of $GP_{Exp}$ decreased. Although the map patterns of $CFSM_{Exp}$ and CFSM-$RF_{Exp}$ differed from those of $GP_{Exp}$, they exhibited better accuracy in uncertainty modeling, as indicated by the lower CRPS values in Table 2 (both under- and over-estimated standard deviation values caused the CRPS to increase).

# 6. Concluding remarks

This study introduces CFSM, a novel spatial modeling framework based on a sequential learning procedure that trains a multiscale spatial process together with linear or non-linear trend functions. Its predictive accuracy and computational efficiency were demonstrated through simulation experiments and empirical applications. Unlike conventional spatial modeling, CFSM optimizes the model via holdout validation, which facilitates seamless integration with machine-learning algorithms, as illustrated with an random forest example. In addition, CFSM is computationally efficient and can be easily parallelized, highlighting its potential as a fast, accurate, and machine-learning-compatible alternative to conventional likelihood-based spatial process models.

Although we demonstrated the predictive accuracy and computational efficiency of CFSM relative to conventional methods, there are other sophisticated spatial learning algorithms, including

GP boost (Sigrist, 2022), geographically weighted forests (GWF; Georganos et al., 2022), multiscale GWF (Cui et al., 2025), and spatial neural networks (e.g, Zhan and Datta, 2025). While the usefulness of CFSM has been demonstrated in this study, future work will explore comparisons with other recently proposed spatial machine learning models.

As an initial step to efficiently combine CFSM with machine learning algorithms, the usefulness of such integration should be examined in terms of both modeling accuracy and interpretation while varying sample sizes and model choices. For instance, integration with a GAM is likely to enhance interpretability, whereas integration with neural networks may improve predictive accuracy in large-scale modeling. These experiments will help in applying the CFSM to other tasks commonly discussed in machine learning literature, such as multitask learning, classification, transfer learning, anomaly detection, and data fusion.

It is also important to extend CFSM itself. Although we assumed spatial processes only for the residuals, these processes have often been used to model spatially varying coefficients (e.g., Gelfand et al., 2003; Comber et al., 2023). Temporal processes, groupwise effects, and other stochastic structures have also been introduced to flexibly model spatiotemporal data (e.g., Wood et al., 2017; Murakami et al., 2025). Extending CFSM to accommodate multiple stochastic processes is an important future research direction. In addition, because CFSM does not rely on the assumption of stationarity as spatial GPs, its efficiency should also be investigated in the context of nonstationary spatial process modeling.

Additionally, loss functions beyond the squared error, such as the Poisson loss for count data and logistic loss for binary outcomes, should be incorporated. Extending the framework in these directions will enable broader applications, including those that remain challenging for conventional spatial statistical models.

The CFSM is implemented in an R package spCF (https://cran.r-project.org/web/packages/spCF/)..

## Acknowledgements

The R code for implementing the linear and non-linear CFSM models for spatial prediction is available at https://github.com/dmuraka/spCL_dev_version/tree/main.

# Appendix 1. Deviations of the local parameter estimates

The local model (Eq. 2) is expressed in matrix form as

$$\mathbf{z}_r = \mu_{c_r}\mathbf{1} + \mathbf{e}_{c_r}, \qquad \mu_{c_r} \sim N\left(0, \tau_{c_r}^2\right), \qquad \mathbf{e}_{c_r} \sim N\left(\mathbf{0}, \sigma_{c_r}^2 \mathbf{W}_{c_r}^{-1}\right), \tag{A1}$$

where $\mathbf{z}_r = [z_r(s_1), \dots, z_r(s_N)]'$, and $\mathbf{1}$ is a vector of ones. $\mathbf{W}_{c_r}$ is a diagonal matrix whose $i$-th element is $w_{h_r}\left(d_{i,c_r}\right)$. Eq. (A1) can be rewritten as

$$\mathbf{z}_{c_r}^{(w)} = \mu_{c_r}\mathbf{w}_{c_r} + \mathbf{e}_{c_r}, \qquad \mu_{c_r} \sim N\left(0, \tau_{c_r}^2\right), \qquad \mathbf{e}_{c_r} \sim N\left(\mathbf{0}, \sigma_{c_r}^2 \mathbf{I}\right), \tag{A2}$$

where $\mathbf{z}_{c_r}^{(w)} = \mathbf{W}_{c_r}^{1/2}\mathbf{z}_r = [w_{h_r}^{1/2}\left(d_{1,c_r}\right)z_r(s_1), \dots, w_{h_r}^{1/2}\left(d_{N,c_r}\right)z_r(s_N)]'$, and $\mathbf{w}_{c_r}^{1/2} = \mathbf{W}_{c_r}^{1/2}\mathbf{1} = [w_{h_r}^{1/2}\left(d_{1,c}\right), \dots, w_{h_r}^{1/2}\left(d_{N,c}\right)]'$. Eq. (A2) is identical to the basic random effects model with response vector $\mathbf{z}_{c_r}^{(w)}$, covariate vector $\mathbf{w}_{c_r}^{1/2}$, and random coefficient $\mu_{c_r}$. As per Wood (2017), the penalized least squares estimate of $\mu_{c_r}$ and its variance are given by

$$\hat{\mu}_c = \left(\left(\mathbf{w}_{c_r}^{1/2}\right)'\mathbf{w}_{c_r}^{1/2} + \frac{\sigma_{c_r}^2}{\tau_{c_r}^2}\right)^{-1} \left(\mathbf{w}_{c_r}^{1/2}\right)'\mathbf{W}_{c_r}^{1/2}\mathbf{z}_{c_r}, \tag{A3}$$

$$Var[\hat{\mu}_c] = \hat{\sigma}_c^2 \left(\left(\mathbf{w}_{c_r}^{1/2}\right)'\mathbf{w}_{c_r}^{1/2} + \frac{\sigma_{c_r}^2}{\tau_{c_r}^2}\right)^{-1}. \tag{A4}$$

It can be confirmed that Eq. (A3) is identical to Eq. (3).

The predictive variance of $\mathbf{z}_{c_r}$ is given by

$$Var\left[\mathbf{z}_{c_r}\right] = Var\left[\mu_{c_r}\mathbf{1} + \mathbf{e}_{c_r}\right] = Var\left[\mu_{c_r}\right]\mathbf{1} + Var\left[\mathbf{e}_{c_r}\right], \tag{A5}$$

$$= \hat{\sigma}_c^2 \left(\left(\mathbf{w}_{c_r}^{1/2}\right)'\mathbf{w}_{c_r}^{1/2} + \frac{\sigma_{c_r}^2}{\tau_{c_r}^2}\right)^{-1} \mathbf{1} + \sigma_{c_r}^2 \mathbf{W}_{c_r}^{-1}.$$

Here, the $i$-th element equals Eq. (5).

## Appendix 2. Predictive mean and variance of the aggregated process

Consider the basic normal distribution $z \sim N(\mu, \sigma^2)$ with a probability density function (PDF) $p(z) = \frac{1}{\sqrt{2\pi\sigma^2}} \exp\left(-\frac{(z-\mu)^2}{2\sigma^2}\right)$. The PDF can be easily expanded as follows:

$$p(z) = \frac{1}{\sqrt{2\pi\sigma^2}} \exp(Az^2 + Bz + C), \quad A = -\frac{1}{2\sigma^2}, \quad B = \frac{\mu}{\sigma^2}, \quad C = -\frac{\mu^2}{2\sigma^2}, \tag{A6}$$

meaning $\sigma^2 = -\frac{1}{2A}$ and $\mu = -\frac{B}{2A}$.

In our case, the PDF of the $c_r$-th local model $z_r(s_i)|\mu_{c_r}, \sigma_{c_r}^2 \sim N(\hat{z}_r(s_i), \sigma_r^2(s_i))$ is

$$p\left(z_r(s_i)\middle|\mu_{c_r}, \sigma_{c_r}^2\right) = \frac{1}{\sqrt{2\pi\sigma_{c_r}^2}} \exp\left(-\frac{\left(z_r(s_i) - \mu_{c_r}\right)^2}{2\sigma_{c_r}^2}\right). \tag{A7}$$

The log of the product of the PDFs of the $C_r$ local models is

$$\prod_{c_r=1}^{C_r} p\left(z_r(s_i)\middle|\mu_{c_r}, \sigma_{c_r}^2\right)^{w_r(d_{i,c_r})} \tag{A8}$$

$$= \prod_{c_r=1}^{C_r} \left(\frac{1}{\sqrt{2\pi\sigma_{c_r}^2}}\right)^{w_r(d_{i,c_r})} \exp\left(-\frac{1}{2}\sum_{c_r=1}^{C_r} \frac{w_r(d_{i,c_r})\left(z_r(s_i) - \mu_{c_r}\right)^2}{\sigma_{c_r}^2}\right),$$

where

$$-\frac{1}{2}\sum_{c_r=1}^{C_r} \frac{w_r(d_{i,c_r})\left(z_r(s_i) - \mu_{c_r}\right)^2}{\sigma_{c_r}^2} = A^* z_r(s_i)^2 + B^* z_r(s_i) + C^*. \tag{A9}$$

Here, $A^* = -\frac{1}{2}\sum_{c_r=1}^{C_r} \frac{w_r(d_{i,c_r})}{\sigma_{c_r}^2}$, $B^* = \sum_{c_r=1}^{C_r} \frac{w_r(d_{i,c_r})\mu_{c_r}}{\sigma_{c_r}^2}$, and $C^* = -\frac{1}{2}\sum_{c_r=1}^{C_r} \frac{w_r(d_{i,c_r})\mu_{c_r}^2}{\sigma_{c_r}^2}$. Based on the relationship between Eqs. (A6) and (A9), we have

$$\sigma^2(s_i) = -\frac{1}{2A^*} = 1/\sum_{c_r=1}^{C_r} \frac{w_r(d_{i,c_r})}{\sigma_{c_r}^2}, \tag{A10}$$

$$\hat{z}(s_i) = -\frac{B}{2A} = \left( \sum_{c_r=1}^{C_r} \frac{w_r(d_{i,c_r})\mu_{c_r}}{\sigma_{c_r}^2} \right) \Big/ \sum_{c_r=1}^{C_r} \frac{w_r(d_{i,c_r})}{\sigma_{c_r}^2}, \tag{A11}$$

which are identical to Eqs. (7) and (8).

## Appendix 3: Learning strategies of the single-scale processes

This appendix compares the predictive accuracy of our linear spatial models (Eq. 11) under different learning strategies for $\boldsymbol{z}_1, \ldots, \boldsymbol{z}_R$. Because some alternatives assume $R$ as known, we fix $R$ = 10. We compare our assumed coarse-to-fine learning with ordering $\boldsymbol{z}_1, \ldots, \boldsymbol{z}_{10}$, with a fine-to-coarse learning with ordering $\boldsymbol{z}_{10}, \ldots, \boldsymbol{z}_1$, as well as the specifications considering only the coarsest-scale process $\boldsymbol{z}_1$ or only the finest-scale process $\boldsymbol{z}_{10}$. All these specifications are assumed in the linear spatial model (Eq. 11), and their predictive accuracy is evaluated in the same manner as in Section 4.1. Figure A1 presents boxplots of the RMSEs over 100 iterations, clearly showing that the coarse-to-fine learning strategy consistently achieves the lowest predictive error.

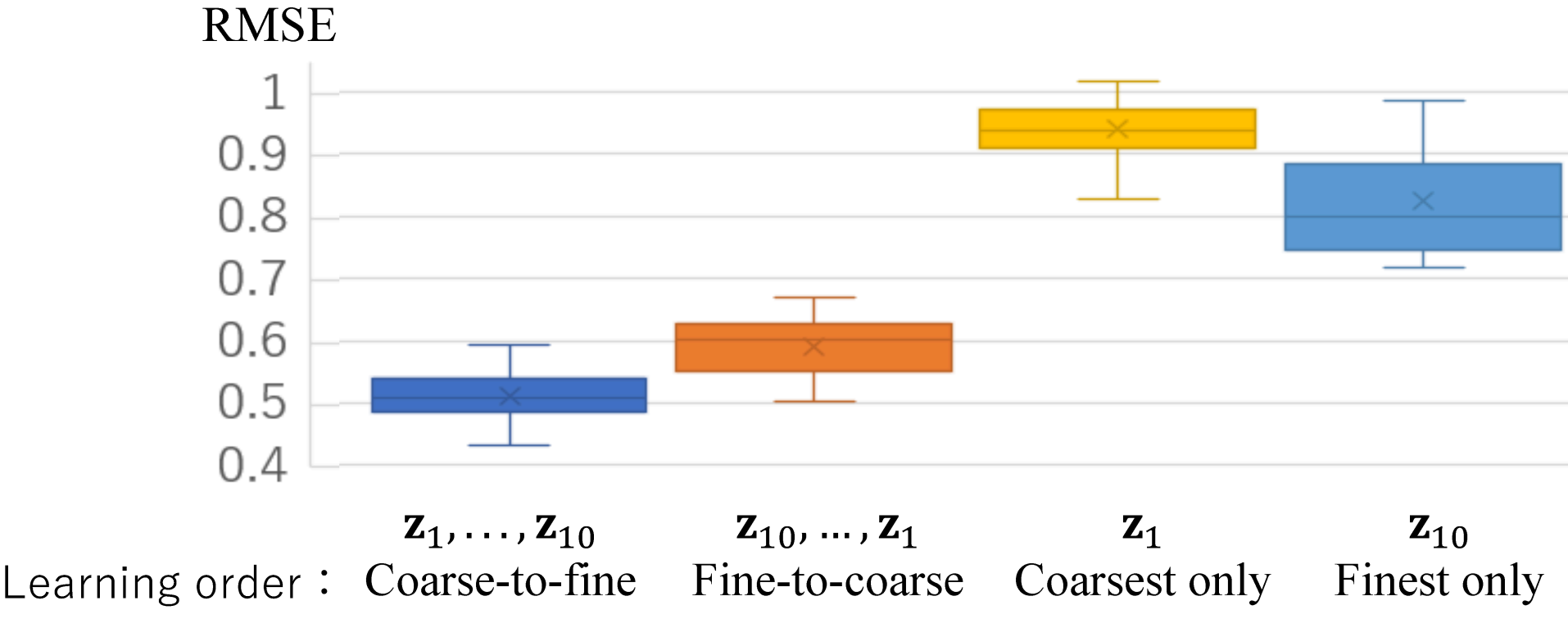

Figure A1. Boxplot of the RMSE values for each learning strategy.

## Appendix 4: Monte Carlo experiments using GPs for data generation

This section presents additional experimental results based on the linear spatial model (Eq. 13), where the spatial moving average processes $z(s_i), z_k(s_i) \in \{z_1(s_i), \ldots,, z_K(s_i)\}$ are replaced with GPs defined as $z(s_i) \sim GP(0, 2^2 k_h(d_{ij}))$ and $z_k(s_i) \sim GP(0, k_1(d_{ij}))$, where $k_h\left(d_{ij}\right) = \exp\left(-d_{ij}^2/h^2\right)$. Owing to computational complexity, experiments could only be conducted for cases with $N \leq 6{,}000$. All other settings followed those described in Section 4.1.

Figure A2 shows the median RMSE and MAE values, which are analogous to those in Figure 8. The results suggest that although NNGP and GP perform best when $h = 0.2$ and $h = 1.0$, respectively, under settings favorable to GP models, CFSM still achieves an accuracy compatible with these best-performing models in each scenario, especially for larger samples.

The poor performance of GP in the case of $h$ = 0.2. is attributable to the property of non-linear weighted least squares estimation that uses the medians of the sample variogram in each distance bin for variogram fitting (Cressie, 2015). In other words, using medians instead of raw samples may mask fine-scale variability.

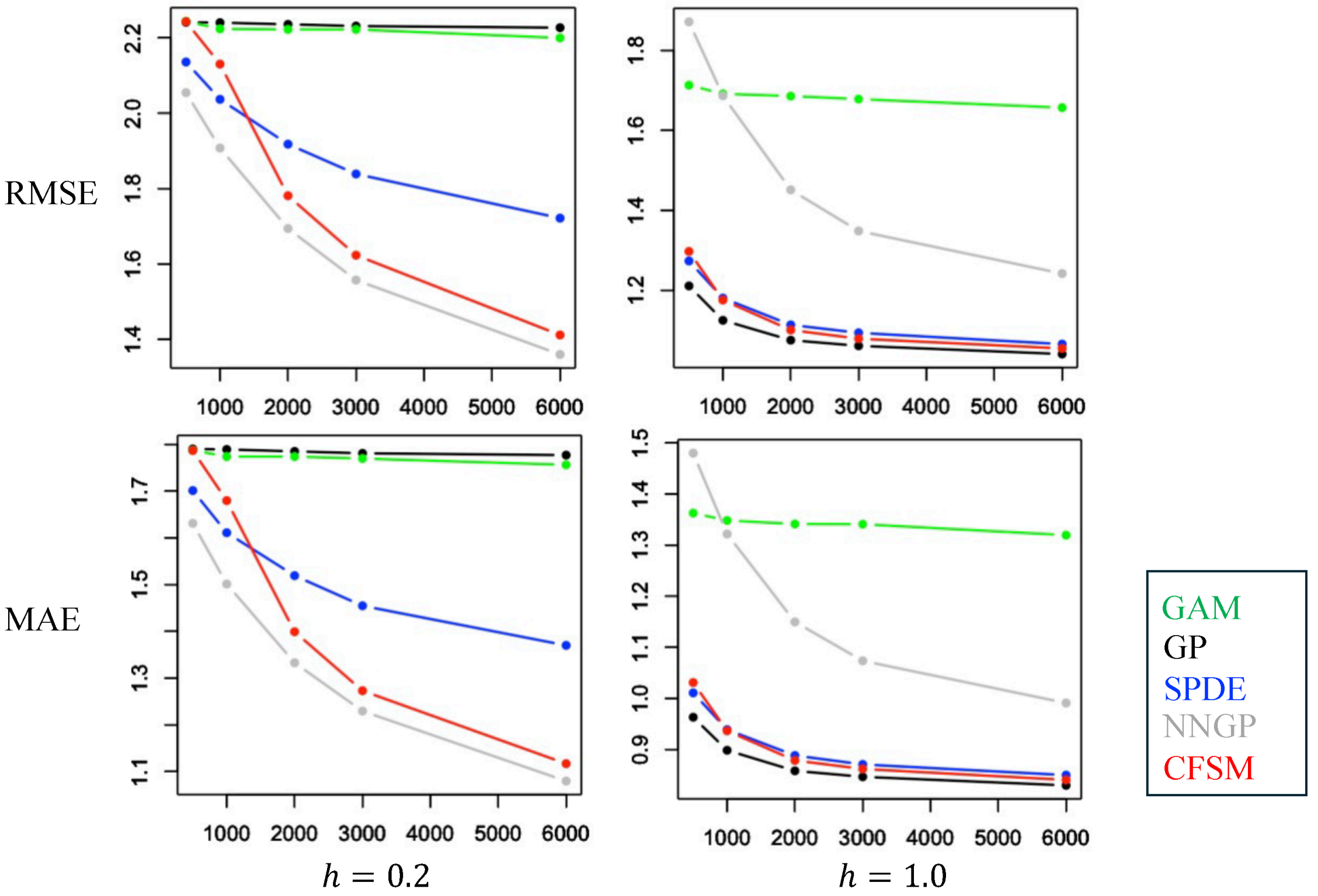

Figure A2. Medians of the RMSE and MAE values (True: GP model) for $N \leq 6{,}000$.